\documentclass[]{article}

\usepackage{amsmath}
\usepackage{amssymb}
\usepackage{tikz}
\usepackage{graphicx}
\usepackage[footnotesize,hang]{caption}
\usepackage{subfig}
\usepackage{color}
\usepackage{rotating}
\usepackage{mathtools}
\usepackage{xcolor}
\usepackage{epstopdf}
\usepackage{cite}
\usepackage{url}

\usepackage[margin=0.85in]{geometry}
\def\Integer{\mathbb{Z}}

\usetikzlibrary{decorations.pathmorphing}

\renewcommand{\vec}[1]{\boldsymbol{#1}}

\newcommand{\eps}{\epsilon}

\renewcommand{\i}{\mathrm{i}}
\newcommand{\e}{\mathrm{e}}

\renewcommand{\d}{\,\mathrm{d}}
\newcommand{\diff}[2]{\frac{\mathrm{d} #1}{\mathrm{d} #2}}

\def\XXint#1#2#3{{\setbox0=\hbox{$#1{#2#3}{\int}$}
\vcenter{\hbox{$#2#3$}}\kern-.5\wd0}}

\title{\textcolor{black}{Nanoptera in weakly nonlinear woodpile chains and diatomic granular chains}}
\author{G. Deng$^1$, C. J. Lustri$^1$\footnote{Corresponding Author. Electronic address: christopher.lustri@mq.edu.au}, and Mason A. Porter$^2$}
\date{%
    $^1$Department of Mathematics and Statistics, 12 Wally's Walk, Macquarie University, New South Wales 2109, Australia\\[2ex]%
    $^2$Department of Mathematics, 520 Portola Plaza, University of California, Los Angeles, California 90024, USA
}                                     

\begin{document}
\maketitle
\abstract{We study ``nanoptera'', which are non-localized traveling waves with exponentially small but non-decaying oscillations, in two singularly-perturbed Hertzian chains with precompression. These two systems are woodpile chains (which we model as systems of Hertzian particles and springs) and diatomic Hertzian chains with alternating masses. We demonstrate that nanoptera arise from Stokes phenomena and appear as special curves, which are called ``Stokes curves'', are crossed in the complex plane. We use techniques from exponential asymptotics to obtain approximations of the oscillation amplitudes.
Our analysis demonstrates that traveling waves in a singularly perturbed woodpile chain have a single Stokes curve, across which oscillations appear. Comparing these asymptotic approximations
with numerical simulations reveals that they accurately describe the non-decaying oscillatory behavior in a woodpile chain. We perform a similar analysis of a diatomic Hertzian chain and show that its nanopteron solution has two distinct exponentially small oscillatory contributions. We demonstrate that there exists a set of mass ratios for which these two contributions cancel to produce localized solitary waves. This result builds on prior experimental and numerical observations that there exist mass ratios that support localized solitary waves in diatomic Hertzian chains without precompression. Comparing asymptotic and numerical results in a diatomic Hertzian chain with precompression reveals that our exponential asymptotic approach accurately predicts the oscillation amplitude for a wide range of system parameters but that it
fails to identify several values of the mass ratio that correspond to localized solitary-wave solutions.

}

%%%%%

\section{Introduction}\label{intro}

%%%%

\subsection{Particle Chains}

The behavior of a particle chain under compression depends on the character of the
interactions
between particles. A key example is
the Hertzian interaction~\cite{Hertz}, which describes the relationship between the repelling force from the contact area of adjacent frictionless spherical particles and the compression between them. \textcolor{black}{Hertzian interactions were first used to describe static chain configurations. Nesterenko~\cite{Nesterenko} demonstrated that it is also reasonable to model dynamical processes such as wave propagation in particle chains using Hertzian interactions between adjacent particles if certain constraints on the particle contact areas, the stresses at the contact points, and the characteristic time scales of the dynamics
are satisfied. These constraints were validated in experimental studies~\cite{Lazaridi,Coste}}.
Chains of particles with Hertzian interactions, which are typically called ``Hertzian chains'', have been the subject of numerous investigations in the last few decades~\cite{Nesterenko1,sen:2008,chong2017} both because they possess rich dynamics and because of their potential uses for practical engineering applications~\cite{porter2015}.

\begin{figure}[tb]
\centering
\includegraphics{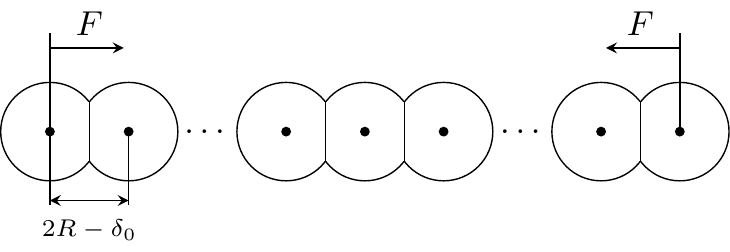}
\caption{Schematic illustration of a particle chain with precompression. The chain consists of $N$ identical, aligned spheres with radius $R$. The interaction potential between adjacent spheres is governed by~\eqref{e:potential}. Applying a compressive force $F$ at both ends of the chain compresses each sphere at the contact point by a uniform distance $\delta_0$. }
\label{f:precompression}
\end{figure}

A general equation that governs the motion of particles in a chain is the following system of differential--difference equations:
\begin{equation}
	m(n)\ddot{x}(n,t)=\phi'(x(n+1,t)-x(n,t))-\phi'(x(n,t)-x(n-1,t))\,,
\label{e:lattice}
\end{equation}
where $n\in\Integer$, the quantity $m(n)$ is the mass of the $n$-th particle, $x(n,t)$ is the position of the $n$-th particle at time $t$, a dot denotes differentiation with respect to time, a prime denotes differentiation with respect to space, and the interaction potential between adjacent particles is
\begin{eqnarray}
	\phi(r)=
\begin{cases}
	c(\delta_0-r)^{\alpha + 1}\,, &r\leq\delta_0 \cr 0\,, &r>\delta_0 \end{cases}\,,~\quad c = \mathrm{constant}\,,
\label{e:potential}
\end{eqnarray}
where $\alpha > 1$ and $\delta_0$ is the equilibrium overlap of adjacent particles that arise from the precompression that is induced by an external force. In Fig.~\ref{f:precompression}, we illustrate a particle chain with precompression. The power-law interaction potential~\eqref{e:potential} is $0$ when particles are not in contact, and it cannot take negative values. For algebraic convenience, we set $c=1/(\alpha+1)$ in all of our examples.
The exponent $\alpha$ in the power-law interaction potential~\eqref{e:potential} depends on the contact geometry between adjacent particles~\cite{Spence1968}.
\textcolor{black}{As was illustrated in~\cite{Fraternali,Sun2011}, $\alpha$ is tunable experimentally. The choice of $\alpha = 3/2$ 
gives the widely-studied Hertzian interaction potential, but it is also relevant to consider other values of the exponent $\alpha$ \cite{chong2017}. For example, Nesterenko~\cite{Nesterenko3} investigated transverse vibrations of unstressed linear elastic fibers, for which $\alpha=3$. Sen and Manciu~\cite{sen2001} and Avalos and Sen~\cite{avalos:2009} investigated chains with power-law interaction potentials for several different values of $\alpha$. The transition from interactions with $\alpha=2$ to Hertzian interactions was investigated in~\cite{Goddard1990}.}

\textcolor{black}{
The existence of solitary waves was first reported for Hertzian chains with precompression in~\cite{Nesterenko}. It was subsequently shown that particle chains with other interaction exponents $\alpha>1$ also support the propagation of solitary waves~\cite{Nesterenko4,Nesterenko5}.} Friesecke and Wattis~\cite{FrieseckeWattis} showed that solitary-wave solutions can occur in nonlinear lattices with a broad class of interaction potentials that are known as ``superquadratic potentials"\footnote{A interaction potential $\phi(r)$ is ``superquadratic'' when $\phi(r)/r^2$ strictly increases with $|r|$ for all $r\in\Lambda$ for either $\Lambda=(-\infty,0)$ or $\Lambda=(0,\infty)$.}, which includes power-law interactions with $\alpha>1$. See~\cite{Nesterenko1,sen:2008,chong2017} for reviews of the properties of solitary waves in granular chains.
\textcolor{black}{Particle chains with $\alpha<1$ do not support solitary waves; see~\cite{Nesterenko6,Yasuda,Herbold2013} for further discussion of such chains.}

Since the first prediction of solitary waves in Hertzian chains~\cite{Nesterenko}, there have been many theoretical, numerical, and experimental studies of the properties of solitary waves in Hertzian chains~{\cite{Nesterenko1,sen:2008,porter2015,chong2017,Nesterenko7}}.
These studies have concerned a wide variety of topics, including the generation~\cite{Deng1,Lazaridi,Coste,DaraioNesterenko2006,Hinch}, propagation~\cite{Deng1,Lazaridi,Coste,DaraioNesterenko2006,Hinch}, interaction~\cite{manciu:2002,job:2005,avalos:2009,avalos:2011,avalos:2014,Deng}, and long-time dynamics \cite{avalos:2011,avalos:2014,sen:2004,Prz:2015,przedborski:2017} of solitary waves in monoatomic Hertzian chains, in which every particle is the same.

The system \eqref{e:lattice}--\eqref{e:potential} that describes the dynamics of a particle chain is both nonlinear and nonintegrable, so typically one cannot study its solutions using exact, fully analytical methods. However, under the assumption of small deformations, such that $|x(n-1,t)-x(n,t)|/\delta_0\ll1$, one can expand the right-hand side of~\eqref{e:lattice} {as a Taylor series in the normalized deformation size~\cite{Nesterenko}.} Neglecting terms of order $O((|x(n-1,t)-x(n,t)|/\delta_0)^4)$ produces the frequently studied Fermi--Pasta--Ulam--Tsingou (FPUT) chain. In the long-wavelength limit, in which the characteristic spatial size of a solution is much larger than the lattice spacing, one can approximate the lattice system \eqref{e:lattice}--\eqref{e:potential}
by a
continuous
system. In the limit of both small deformations and long wavelengths 
{in comparison to particle size}, the monoatomic Hertzian system with precompression reduces to the Kortweg--de Vries (KdV) equation \cite{Nesterenko1}. In this long-wavelength limit, one can approximate the solitary-wave behavior in a Hertzian chain
as a soliton solution of the KdV equation \cite{kdv}. Building on this idea, the interaction of solitary waves in a Hertzian chain in the long-wavelength limit was described in~\cite{Shen} using a two-soliton solution of the KdV equation.

\begin{figure}
\centering
\subfloat[A physical woodpile system]{
\includegraphics{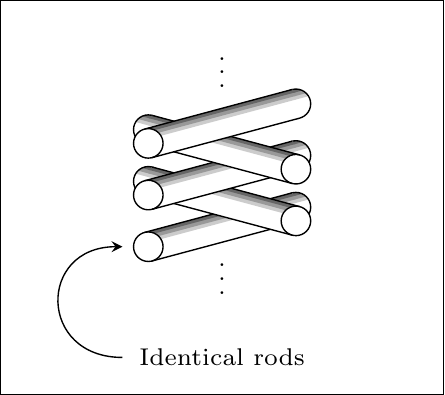}
}
\subfloat[A woodpile-chain model]{
\includegraphics{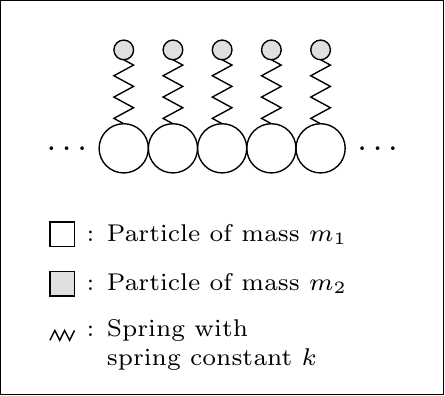}
}
\subfloat[A diatomic chain of particles]{
\includegraphics{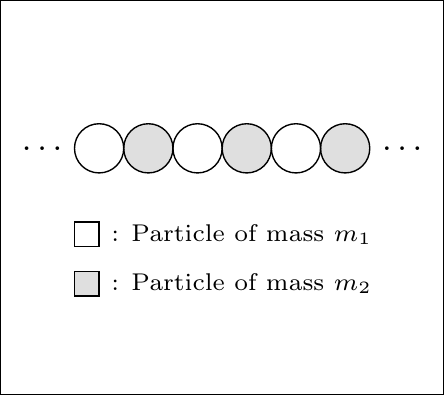}
}
\caption{The chains of particles that we examine in the present study.
The schematic illustration in (a) shows the physical configuration of orthogonal rods that is known as a ``woodpile chain''. The schematic in (b) shows an idealized mathematical model of the physical configuration in (a). This model consists of heavy spherical particles in physical contact via interactions governed by~\eqref{e:potential} and light spherical particles (which are sometimes called ``resonators'') that are attached to each heavy particle by a spring. The schematic illustration in (c) shows a diatomic chain of alternating heavy and light spheres in physical contact via Hertzian interactions.
We analyze the models in (b) and (c).
}
\label{f:DiatomicWoodpile}
\end{figure}

A remarkable feature of Hertzian chains (and higher-dimensional generalizations of them) is their tunability. It is possible to produce a wide range of dynamical behaviors through simple modifications of a homogeneous Hertzian system, such as by incorporating different types of particle heterogeneities. Such modifications can produce behavior that is very different from that of homogeneous Hertzian chains. For example, researchers have explored the behavior of Hertzian chains with impurities~\cite{SenManciuWright,Hascoet,Martinez1}, \textcolor{black}{disordered
particle arrangements~\cite{KimMartinez,Martinez2,sen:2000,Harbola,Manjunath,Nesterenko},} quasiperiodicity~\cite{Martinez3}, and particles that are composed of segments with different materials (i.e., so-called ``compound chains'')~\cite{Vergara2005,Vergara2006,NestrenkoDaraio2005,DaraioNesterenko2006_1}.
It has been demonstrated that impurities can cause solitary waves to scatter~\cite{SenManciuWright,Hascoet,Martinez1}. Transport and localization of energy in Hertzian systems with disorder and quasiperiodicity were studied in~\cite{KimMartinez,Martinez2,Martinez3}.
In Hertzian chains with randomly arranged masses, solitary waves can delocalize, such that the energy that is carried by a solitary wave spreads among many particles in a chain~\cite{sen:2000,Harbola,Manjunath}.
The scattering of solitary waves at an interface of a compound Hertzian system was studied in~\cite{Vergara2005,NestrenkoDaraio2005}. Other phenomena that have been studied in numerical and experimental investigations of heterogeneous Hertzian chains include energy trapping~\cite{Vergara2006,DaraioNesterenko2006_1}, shock disintegration~\cite{DaraioNesterenko2006_1}, and the generation of secondary solitary waves~\cite{Vergara2005}.

It is also interesting to consider other types of heterogeneous chains, including (1) particle--rod systems that model so-called ``woodpile chains''~\cite{chong2017,Kim,Xu,Liu2,Liu1,Jiang,Wu,KimYang} and (2) diatomic chains~\cite{chong2017,Theocharis,Boechler2010,Molinari,Ponson,Porter2008,Porter2009,Hoogeboom,Herbold2009,Jayaprakash,Jayaprakash1,KimChaunsali2015,Potekin2013}. It is feasible to study both woodpile chains and diatomic chains in laboratory experiments.

A woodpile chain consists of orthogonally stacked slender rigid cylinders with mass $m_1$ [see Fig.~\ref{f:DiatomicWoodpile}(a)]. The interaction force along the direction of the stack is determined by~\eqref{e:potential}, and one can model the elastic deformation along the direction that is perpendicular to the stack direction using internal ``resonators'', where the mass and coupling constant of these resonators come from
the material and shape of the cylinders. When considering a stack of identical cylinders, each resonator has the same mass $m_2$ and elastic constant $k$. We model such a woodpile chain as a monoatomic granular chain in which each particle
is connected to an external particle by a rod, which we treat as a linear spring. Each particle in the chain has mass $m_1$, each external particle has mass $m_2$, and each linear spring has spring constant $k$. We illustrate this woodpile chain model in Fig.~\ref{f:DiatomicWoodpile}(b). For the rest of the present study, we use the term ``woodpile chain'' to refer to this idealized system.
Much of the existing work on woodpile chains has concentrated on their linear elastic responses \cite{Jiang,Wu,KimYang}. Some recent work studied the behavior of traveling waves in woodpile chains with no precompression~\cite{Kim,Xu} and hence in a strongly nonlinear regime.
Other studies have investigated the existence of discrete breathers in woodpile chains both without precompression~\cite{Liu2} and with precompression~\cite{Liu1}.

A diatomic chain consists of particles with some characteristic, such as particle mass, that alternates in adjacent particles. We consider chains [see Fig.~\ref{f:DiatomicWoodpile}(c)] in which even particles have mass $m_1$ and odd particles have mass $m_2$. Diatomic Hertzian chains have rich dynamics~\cite{chong2017}. They can have both discrete breather solutions~\cite{Theocharis,Boechler2010} and stationary shock waves~\cite{Molinari}. Previous studies have investigated the propagation and scattering of nonlinear waves in both ordered and disordered diatomic Hertzian chains~\cite{Ponson,Porter2008,Porter2009,Herbold2009} and chaotic dynamics in diatomic Hertzian chains that are
damped and driven~\cite{Hoogeboom}.

%%%%%

\subsection{Nanoptera}

\begin{figure}
\centering
\subfloat[Solitary wave]{
%\begin{tikzpicture}
%[xscale=0.4,>=stealth,yscale=2]
%\draw[black,thick] plot[smooth] file {Sol1b.txt};
%\draw (-5,1.2) -- (5,1.2) -- (5,-0.2) -- (-5,-0.2) -- cycle;
%\draw [->] (1,0.5) -- (2,0.5) node[above right] {\scriptsize{$x = c_0t$}};
%\draw node at (-6,0.55) [rotate=90] {\scriptsize{Displacement}};
%\draw node at (0,-0.3) {\scriptsize{$x$}};
%\end{tikzpicture}
\includegraphics{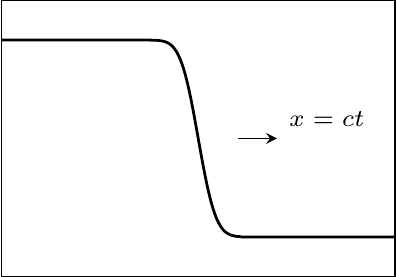}
}
\subfloat[One-sided nanopteron]{
%\begin{tikzpicture}
%[xscale=0.4,>=stealth,yscale=2]
%\draw[black,thick] plot[smooth] file {Sol2b.txt};
%\draw (-5,1.2) -- (5,1.2) -- (5,-0.2) -- (-5,-0.2) -- cycle;
%\draw [->] (1,0.5) -- (2,0.5) node[above right] {\scriptsize{$x = c_1t$}};
%\draw node at (0,-0.3) {\scriptsize{$x$}};
%\end{tikzpicture}
\includegraphics{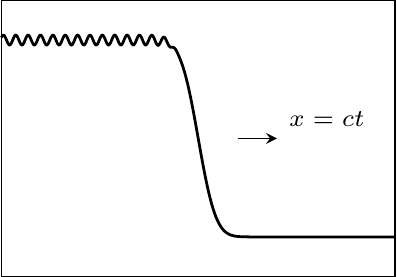}
}
\subfloat[Two-sided nanopteron]{
%\begin{tikzpicture}
%[xscale=0.4,>=stealth,yscale=2]
%\draw[black,thick] plot[smooth] file {Sol3b.txt};
%\draw (-5,1.2) -- (5,1.2) -- (5,-0.2) -- (-5,-0.2) -- cycle;
%\draw [->] (1,0.5) -- (2,0.5) node[above right] {\scriptsize{$x =c_2t$}};
%\draw node at (0,-0.3) {\scriptsize{$x$}};
%\end{tikzpicture}
\includegraphics{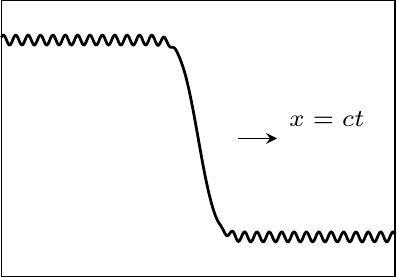}
}
\caption{Comparison of the profiles of (a) a standard solitary wave, (b) a one-sided nanopteron, and (c) a two-sided nanopteron. \textcolor{black}{The waves propagate at speeds of $c_0$, $c_1$, and $c_2$ respectively.} The solitary wave is localized spatially, whereas the nanoptera have non-decaying oscillatory tails on (b) one side or (c) both sides of the wave front. The waves in (a) and (c) propagate without decaying, but the wave in (b) cannot propagate indefinitely.
\textcolor{black}{In (b), the nanoptera decay \textcolor{black}{very slowly} 
{because}
the one-sided radiation draws energy from the wave front.}
{Additionally, because this decay occurs over a long time scale, the one-sided nanopteron in (b) is said to be ``metastable''.}
}
\label{f:nanoptera}
\end{figure}

Typically, traveling-wave solutions in woodpile and diatomic Hertzian systems are not truly localized~\cite{Jayaprakash,Kim,Xu}; instead, they also include a small non-localized oscillatory tail and thus take the form of a nanopteron~\cite{Boyd}. A nanopteron is the superposition of a central solitary wave and a persistent oscillation in the ``far field'' (which refers to the region in which the distance from the central solitary wave tends to infinity) on one or both sides of the central wave\footnote{Several studies have used the term ``nanopteron'' to refer only to the additional small oscillations. We do not follow this convention; instead, we use the term ``nanopteron'' to refer to a solution that includes both a central wave and such small oscillations.}.
 These oscillations typically have an exponentially small amplitude with respect to \textcolor{black}{a singular perturbation parameter. (In the present study, this parameter is the mass ratio between one type of particle and the other type of particle.)}
 We show examples of nanoptera in Fig.~\ref{f:nanoptera}. A genuine solitary wave is exponentially localized, but a nanopteron is localized only up to algebraic orders in the small parameter.

\textcolor{black}{Jayaprakash \emph{et al.}~\cite{Jayaprakash} showed that for uncompressed diatomic Hertzian chains non-decaying oscillations are absent at certain values of the system parameters}. These sets of parameters arise from satisfying
an ``anti-resonance condition". When such anti-resonance condition is satisfied,  nanoptera in uncompressed diatomic Hertzian chains become solitary waves that propagate without attenuation. Other studies that have examined anti-resonance conditions in diatomic Hertzian chains include~\cite{KimChaunsali2015,Potekin2013}.
Hertzian chains can also have
nanopteron solutions that travel with particularly strong dispersion and attenuation.
Such behavior occurs if a chain satisfies
a ``resonance condition''~\cite{Jayaprakash1,KimChaunsali2015,Potekin2013}, which one obtains in a similar fashion to an anti-resonance condition. The study of {nanopteron solutions} in one-dimensional Hertzian chains~\cite{Jayaprakash} was extended to two-dimensional {Hertzian} systems by Manjunath \emph{et al.}~\cite{Manjunath2014}, who obtained anti-resonance conditions for these systems and examined the properties of solitary-wave solutions when these anti-resonance conditions are satisfied. Anti-resonance conditions were also identified for uncompressed woodpile chains by Xu et al.~\cite{Xu}.

Nanoptera have been studied in many other particle chains, such as diatomic Toda chains~\cite{Vainchtein,Okada,Tabata,Lustri}, diatomic FPUT chains~\cite{Vainchtein,Faver,Hoffman,Lustri1}, and chains with an on-site nonlinear potential
in which each particle is linearly coupled to its neighbors~\cite{Iooss}. The existence of nanopteron solutions in a diatomic FPUT chain was proven rigorously in~\cite{Faver,Hoffman}, and nanoptera in periodic Toda chains were studied using a multiple-scale approach in~\cite{Vainchtein}. Recently,~\cite{Faver2} studied diatomic FPUT chains using numerical continuation. They explored the relation between nanopteron solutions for chains in which one mass is much larger than the other (i.e., the so-called ``small-mass-ratio'' regime) and
traveling-wave solutions {for chains with a mass ratio close to $1$.} 

Anti-resonance conditions, at which the small oscillations disappear entirely and a nanopteron becomes a genuine solitary wave, have been identified in diatomic Toda chains~\cite{Vainchtein,Tabata,Lustri}, diatomic FPUT chains~\cite{Lustri}, woodpile chains without
precompression~\cite{Xu}, and a discrete nonlinear Schr\"{o}dinger equation~\cite{Alfimov}.

An exponential asymptotic method was applied recently by Lustri \emph{et al.}~\cite{Lustri,Lustri1} to study exponentially small oscillations in diatomic Toda and FPUT chains.
They showed that the tailing oscillations in these diatomic systems are examples of the ``Stokes phenomenon'', which refers to behavior that is switched on when special curves (called ``Stokes curves'') are crossed in the complex plane.
They demonstrated that traveling-wave solutions in these diatomic chains possess two Stokes curves, which generate two distinct oscillations with the same amplitude but different phases. {For particular values of the mass ratio, the oscillations are precisely out of phase. The oscillations thus cancel, and the solution is a localized solitary wave.}
In \cite{Lustri,Lustri1}, Lustri \emph{et al.}~also derived asymptotic anti-resonance conditions for mass ratios
for these systems. In the present study, we use exponential asymptotic methods to examine nanoptera in singularly perturbed woodpile and diatomic Hertzian chains under precompression.

As we illustrated in Fig.~\ref{f:nanoptera}, nanoptera can have oscillations on either one or both sides of a central traveling wave. The existence of symmetric non-decaying two-sided nanopteron solutions has been proven for {a} fifth-order KdV equation~\cite{Hunter1988}.
Subsequently, symmetric nanopteron solutions of this fifth-order KdV equation were constructed numerically using perturbation series in Boyd~\cite{Boyd1}; Boyd then showed that his numerical approach cannot be used
to compute one-sided nanopteron solutions in that system.
It was proven in~\cite{Benilov} that one-sided nanopteron solutions for the fifth-order KdV equation cannot
{propagate indefinitely without changing form}. The one-sided oscillation draws energy from the central wave, which causes the amplitude of the central core to decay.
{The eventual decay of nanopteron solutions of the fifth-order KdV equation is a result of energy conservation,
and one can make a similar argument against the existence of one-sided nanoptera in any energy-conserving system.}
The instability of one-sided nanopteron solutions in the fifth-order KdV equation was discussed further in~\cite{Grimshaw}.

Giardetti \emph{et al.}~\cite{Giardetti} simulated one-sided nanoptera in a diatomic FPUT model with small mass ratios for a long enough time to obtain a solution that appeared to be a stable traveling wave. In their simulations, both the central solitary wave and the oscillatory tail traveled without any apparent decay. However, {the energy of the system that they studied is conserved.}
Consequently, it is impossible for both the central core and the one-sided oscillation to have a constant amplitude indefinitely because the one-sided oscillation constantly draws energy from the central core~\cite{Giardetti,Jayaprakash,Vainchtein}. Therefore, it was conjectured in~\cite{Giardetti} that the decay of one-sided nanoptera in diatomic FPUT chains occurs on a time scale that is
{exponentially large in the limit that the mass ratio parameter tends to $0$.}
{Solutions that decay on exponentially slow time scales are sometimes known as
``metastable" solutions (they are also sometimes called ``quasistable" solutions), which describe asymptotic solutions that appear to be stable over a duration that is large in comparison to any inverse power of a small
parameter before eventually decaying.}
In experimental settings, nanoptera are typically generated by a pulse moving through an undisturbed chain. This produces one-sided nanoptera, with no oscillations in the undisturbed region ahead of the pulse. To the best of our knowledge, there do not currently exist experimental realizations of two-sided nanoptera. 
The nanoptera that we consider
are all one-sided, and therefore metastable; however, {as we will note at the end of Section~\ref{s:woodstokes}},
it is straightforward to extend our analysis to the stable case of two-sided nanoptera, which is of theoretical interest.

In the present study, we compute the behavior of nanopteron solutions in precompressed woodpile and diatomic Hertzian chains in the asymptotic limit $m_2/m_1\to0$. In this limit, both the woodpile and the diatomic Hertzian chains are singularly perturbed around a monoatomic Hertzian chain. Jayaprakash \emph{et al.}~\cite{Jayaprakash} investigated this limit for uncompressed diatomic Hertzian chains, but existing studies of woodpile chains have not focused on the small-mass-ratio regime.

In the present paper, we apply an exponential asymptotic method that was developed in~\cite{Chapman,Daalhuis} and was used in \cite{Lustri,Lustri1} to study traveling waves in diatomic Toda and FPUT chains with small mass ratios. In a typical asymptotic power-series analysis, one expands a solution as an algebraic series in some small parameter.
 One then computes the terms in the series using a recursion relation that one obtains by asymptotic matching. This process can never capture exponentially small behavior, which is smaller than any algebraic series term in the asymptotic limit. By contrast, the exponential asymptotic method that we use in the present paper is capable of describing asymptotic behavior on this exponentially small scale. Additionally, it only requires the direct computation of the leading-order series behavior. We describe this exponential asymptotic approach in detail in Section \ref{s:exponential}.

%%%%

%%%%

\subsection{Outline of the Paper}

We first investigate traveling-wave solutions in a woodpile chain with precompression in the limit that
{the wavelength of the central traveling wave is large in comparison to the particle size; this is a weakly nonlinear regime}. We show that the traveling waves in this system are nanoptera and that the appearance of non-decaying exponentially small oscillations behind the wave front arises because of
the Stokes phenomenon.
Using an exponential asymptotic approach, we obtain simple asymptotic expressions for the exponentially small, constant-amplitude waves in the wake of the leading-order solitary wave.
We demonstrate that the solution has a single oscillatory contribution in the wake of the leading wave and thus that it can never be canceled out. Consequently, the woodpile chains that we examine do not have an anti-resonance condition. This contrasts with the results of Xu \emph{et al.}~\cite{Xu}, who identified anti-resonance conditions for a woodpile system with no precompression.

We then study traveling waves in a diatomic Hertzian chain using the same exponential asymptotic approach. The key difference between diatomic Hertzian chains and woodpile chains is that traveling waves in diatomic Hertzian chains have two Stokes curves; this yields solutions with two oscillatory contributions. For certain mass ratios, the two oscillatory contributions cancel precisely and the traveling-wave solution becomes a genuine localized solitary wave.

For both woodpile chains and diatomic Hertzian chains, we use the long-wavelength approximation from \cite{Nesterenko1} to describe the leading-order solitary wave behavior (as in \cite{Lustri1}). This differs from a diatomic Toda system~\cite{Lustri}, for which there exists an analytical expression for the leading-order wave behavior. Therefore, our analysis of woodpile and diatomic Hertzian chains has two distinct small parameters, which correspond to the small mass ratio and the large traveling-wave length scale.

Our paper proceeds as follows. In Section~\ref{s:exponential}, we discuss the exponential asymptotic method that we employ for our analysis.
In Sections~\ref{s:woodLO}--\ref{s:woodstokes}, we use this exponential asymptotic method to obtain an asymptotic approximation of nanopteron solutions in a singularly perturbed woodpile chain. In Section~\ref{s:woodnum}, we compare the results of this approximation with numerical computations. We obtain strong agreement between the asymptotic and numerical results. In Sections~\ref{s:leading}--\ref{s:stokes}, we perform a similar exponential asymptotic analysis on a singularly perturbed diatomic Hertzian chain. In Section~\ref{s:dimernum}, we compare our asymptotic and numerical results for the diatomic Hertzian chain. From this comparison, we see that the employed exponential asymptotic method is useful for approximating the solution behavior for a wide range of mass ratios. However, it fails to detect some important features that arise from the interference between oscillatory contributions. In Section~\ref{s:conclusion}, we conclude and further discuss our results.

%%%%%%

\section{Exponential Asymptotics}
\label{s:exponential}

We aim to determine the asymptotic behavior of the exponentially small oscillations in the wake of the solitary-wave front in singularly perturbed variants of a Hertzian chain. We use $\eta$ to denote the associated small parameter, which is related to mass ratio and satisfies $\eta^2=m_2/m_1$.
It is impossible to determine the form of these oscillations using only classical asymptotic power series because the oscillation amplitude is exponentially small in the asymptotic limit $\eta\to0$. In particular, the amplitude is smaller than any algebraic power of $\eta$. Therefore, we use an exponential asymptotic approach to study the exponentially small oscillations.

{Consider a singularly perturbed differential equation of the form
\begin{equation}
	F(x,g(x),g'(x),g''(x), \ldots;\eta)=0\,,
\label{e:intro_equation}
\end{equation}
where $\eta$ is a small
parameter. To find an asymptotic solution of~\eqref{e:intro_equation}, we typically
expand a solution about some leading-order approximation as a power series in the perturbation parameter $\eta$. Once we have obtained this leading-order solution, we analytically continue it into the complex plane. Typically, this solution is singular at some set of points, which are the end points of Stokes curves~\cite{Stokes}. }Stokes curves play an important role in the present study: as a Stokes curve is crossed, the amplitude of the exponentially small contribution undergoes a smooth but rapid change in value. In the present work, we require the exponentially small contributions to vanish on one side of the Stokes curve, so they ``switch on'' as the Stokes curves are crossed.

In general, an asymptotic power series of the solution of a singularly perturbed {system} is divergent~\cite{Dingle}. Nevertheless, one can truncate a divergent asymptotic series to approximate the
solution~\cite{Boyd2005}. If one chooses the truncation point to minimize the difference between the approximation and the exact solution, the approximation error is typically exponentially small in the asymptotic limit~\cite{Boyd1999}. This is called ``optimal truncation''; we use $N_{\mathrm{opt}}$ to denote an optimal truncation point. We express the solution of a singularly perturbed {system} as the sum of an optimally truncated power series and an exponentially small contribution.
{Substituting this sum into
\eqref{e:intro_equation} produces an equation that describes the behavior of the exponentially small contribution.}

Early work that applied this idea includes~\cite{Berry1988,Berry}, which determined the Stokes switching behavior in several important special functions. The analysis in~\cite{Berry} demonstrated that the switching behavior depends predictably on the manner in which the asymptotic power series diverges.
Subsequent work in~\cite{BerryHowls1990} established techniques known as ``hyperasymptotics" to further reduce the exponentially small error that is generated by truncating 
asymptotic power series.
See~\cite{Berry1991} for a summary and discussion of the results in~\cite{Berry,BerryHowls1990}.

In the present study, we apply an exponential asymptotic method that was developed in~\cite{Chapman,Daalhuis}. We express the solution $g$ of the governing equation as an asymptotic power series
\begin{equation}\label{e:series_intro}
	g \sim \sum_{j=0}^\infty \eta^{r j}g_j \quad \mathrm{as} \quad \eta \rightarrow 0\,,
\end{equation}
where $r$ is the number of times that one needs to differentiate $g_{j-1}$ to obtain $g_j$.

We first substitute the series \eqref{e:series_intro} into the governing equations~\eqref{e:intro_equation}, and we then asymptotically match terms in the subsequent expression to obtain a recursion relation for $g_j$. In a singularly perturbed problem, obtaining $g_j$ using the recursion relation typically requires differentiating earlier terms in the series.
If the series terms have singular points, this repeated differentiation ensures that the series terms diverge in a predictable fashion that is called a ``factorial-over-power divergence''~\cite{Dingle}. As $j$ becomes large, behavior of this form dominates the series terms.

To capture
the factorial-over-power divergence, we write an ansatz for the behavior of $g_j$ in the limit $j \rightarrow \infty$. This yields the so-called ``late-order terms'' of an asymptotic series. As $j \rightarrow \infty$, factorial-over-power divergent behavior dominates the series terms. Motivated by this pattern, Chapman \textit{et. al}~\cite{Chapman} proposed applying a late-order ansatz to approximate the form of the late-order terms, even if computing earlier terms in a series is challenging or intractable. This ansatz is a sum of terms of the form
\begin{equation}
	g_j\sim\frac{G\Gamma(r j+\gamma)}{\chi^{r j+\gamma}}\quad \mathrm{as} \quad j\to\infty\,,
\label{e:lateorder_intro}
\end{equation}
where the parameter $\gamma$ is constant and $G$ and $\chi$ are functions of any
variables but are independent of $j$. The function $\chi$ is known as the ``singulant" and it is equal to $0$ at one or more singularities of the leading-order solution $g_0$. This ensures that the late-order ansatz for $g_j$ is singular at the same {location(s)} as the leading order, with a singularity strength that grows as $j$ increases.
Substituting~\eqref{e:lateorder_intro} into the differential equation~\eqref{e:intro_equation} and matching orders of $j$ allows us to determine the functional forms of $\chi$ and $G$. We determine $\gamma$ by requiring that the late-order behavior is consistent with the local behavior of the leading order in the neighborhood of singular points.

We optimally truncate the power series~\eqref{e:series_intro} using the additional information from the late-order ansatz \eqref{e:lateorder_intro}. Optimal truncation points typically occur after an asymptotically large number of terms, so we apply the late-order ansatz~\eqref{e:lateorder_intro} to obtain a simplified expression
using the heuristic that was described in~\cite{Boyd1999}. This heuristic requires truncating the series after the value of $N_{\mathrm{opt}}$ for which the term $\eta^{N_{\mathrm{opt}}} g_{N_{\mathrm{opt}}}$ has the smallest magnitude.

We use the truncated series to approximate the exact solution with exponentially small error. We write
\begin{equation}
	g = \sum_{j=0}^{N_{\mathrm{opt}}-1} \eta^{r j}g_j+g_{\exp}\,,
\label{e:series_intro_1}
\end{equation}
where $g_{\exp}$ denotes the exponentially small error term and $N_{\mathrm{opt}}$ is the optimal truncation point that we calculate using
the late-order ansatz \eqref{e:lateorder_intro}.

Substituting the truncated series expression \eqref{e:series_intro_1} into the differential equation \eqref{e:intro_equation} produces an equation for the exponentially small remainder term~\cite{Daalhuis}. Away from Stokes curves, one can determine this remainder using a straightforward application of the WKB method \cite{hinch1991}. In the neighborhood of the Stokes curves, we apply an exponential ansatz for the exponentially small term $g_{\exp}$ and write
\begin{equation}
	g_{\exp}\sim\mathcal{S}G\e^{-\chi/\eta} \quad \mathrm{as} \quad \eta \rightarrow 0\,,
\end{equation}
where $\mathcal{S}$ is a function that is known as the ``Stokes multiplier". The Stokes multiplier is approximately constant except in the neighborhood of Stokes curves, where it varies rapidly in a neighborhood of width $\mathcal{O}(\sqrt{\eta})$ as $\eta\to0$. This behavior is known as ``Stokes switching". The locations of 
Stokes curves are determined completely by the form of the singulant $\chi$. As was shown in~\cite{BerryHowls1990}, Stokes curves occur only in locations where $\chi$ is real and positive. We use exponential asymptotics to directly calculate the exponentially small contributions to the asymptotic behavior of solutions that appear as the Stokes curves are crossed. One cannot express these contributions using classical asymptotic power series.

{It is often the case that this exponential asymptotic approach requires the explicit calculation of only the leading-order solution in \eqref{e:series_intro} to determine the exponentially small contributions}.
This is convenient because it can be complicated or even intractable to compute series terms beyond a leading-order expression in nonlinear problems. See~\cite{Boyd1998,Boyd1999} for more details on exponential asymptotics and their applications to nonlocal solitary waves,~\cite{Berry,Berry1991} for examples of other studies of exponential asymptotics, and~\cite{Chapman,Daalhuis} for more details about the particular methodology that we apply in the present paper.

%%%

\section{A Singularly Perturbed Woodpile Chain}
\label{s:woodpile}

We consider a precompressed woodpile chain, such as the one in Fig.~\ref{f:DiatomicWoodpile}(a). We model it by a particle chain; each particle has mass $m_1$ and is connected to an outside particle of mass $m_2$ by a linear spring with spring constant $k$. The governing equations of this idealized model, which we illustrated in Fig.~\ref{f:DiatomicWoodpile}(b), are
\begin{align}\label{1:gov0a}
	m_1 \ddot{u}(n,t) &= [\delta_0  + u(n-1,t) - u(n,t)]_+^\alpha - [\delta_0  + u(n,t) - u(n+1,t)]_+^\alpha - k [u(n,t) - v(n,t)]\,,\\
	m_2 \ddot{v}(n,t) &= k [u(n,t) - v(n,t)]\,,
	\label{1:gov0b}
\end{align}
where $\delta_0$ is the precompression and $u(n,t)$ and $v(n,t)$, respectively, denote the displacements of the $n$-th particle of mass $m_1$ and $m_2$ at time $t$. The choice $\alpha = 3/2$ corresponds to a classical Hertzian chain. Our analysis in this section is valid for any choice of $\alpha$ for which one can approximate the leading-order solution
by a soliton solution of the KdV equation. In practice, this corresponds to $\alpha > 1$ \cite{FrieseckeWattis}.
The subscript $+$ indicates that we evaluate the bracketed term only if its argument is positive; it is equal to $0$ when its argument is negative. That is, we only 
have a Hertzian interaction
between particles that are in physical contact with each other.

We scale the system \eqref{1:gov0a}--\eqref{1:gov0b} using $u  = m_1^{1/(\alpha-1)} \hat{u}$ and $v= m_1^{1/(\alpha-1)} \hat{v}$, and we rewrite the governing equations in terms of $\hat{\delta} = \delta_0/m_1^{1/(\alpha-1)}$ and $\hat{k} = k/m_1$. Setting $\eta^2 = m_2/m_1$ gives the following scaled governing equations:
\begin{align}\label{1:gov1}
	\ddot{\hat{u}}(n,t) &= [\hat{\delta}  + \hat{u}(n-1,t) - \hat{u}(n,t)]_+^\alpha - [\hat{\delta}  + \hat{u}(n,t) - \hat{u}(n+1,t)]_+^\alpha - \hat{k} [\hat{u}(n,t) - \hat{v}(n,t)]\,,\\
\eta^2 \ddot{\hat{v}}(n,t) &= \hat{k} [\hat{u}(n,t) - \hat{v}(n,t)]\,.
\label{1:gov2}
\end{align}
In our study, we assume that the heavy particles always maintain physical contact with neighboring particles.
Therefore, the argument inside a bracket is always non-negative, and we omit the subscript $+$ in our subsequent analysis. One can validate this assumption by checking the solution behavior directly.
For convenience, we perform our analysis on the scaled system \eqref{1:gov1}--\eqref{1:gov2}; in our subsequent notation, we omit the hats that indicate this scaling.

We are interested in the asymptotic behavior of traveling-wave solutions of \eqref{1:gov1}--\eqref{1:gov2} when $0 < \eta \ll 1$. These solutions consist of a localized wave core, which we generate using an asymptotic power series in $\eta$, and exponentially small but non-decaying oscillations that we compute using exponential asymptotic techniques.

%%%%

\subsection{Leading-Order Solution}\label{s:woodLO}

We expand $u(x,t)$ and $v(x,t)$ as an asymptotic power series in $\eta^2$ in the limit $\eta \rightarrow 0$ and write
\begin{align}
	u(n,t)\sim\sum_{j=0}^\infty \eta^{2j}u_j(n,t)\,, \quad v(n,t)\sim\sum_{j=0}^\infty \eta^{2j}v_j(n,t)\,.
\label{1:asympseries}
\end{align}

To find a leading-order solitary wave, which is the first step to construct a nanopteron solution of \eqref{1:gov1}--\eqref{1:gov2},
we apply the series expression~\eqref{1:asympseries} to~\eqref{1:gov2} and match at leading order in the limit $\eta\to0$ to obtain
\begin{align}
	u_0(n,t) = v_0(n,t)\,.
\label{1:leading}
\end{align}
Inserting the relation~\eqref{1:leading} into
\eqref{1:gov1} gives
\begin{align}
	\ddot u_0(n,t)=\left[\delta+u_0({n-1},t)-u_0({n},t)\right]^\alpha-\left[\delta +u_0({n},t)-u_0({n+1},t)\right]^\alpha\,.
\label{1:woodpile12}
\end{align}
One can obtain an approximation to the leading-order behavior by following the analysis in~\cite{Nesterenko1}. 
Assume that the deformation in the chain is sufficiently small such that $|u_{0}(n-1,t) - u_{0}(n,t)| \ll \delta$. We expand~\eqref{1:woodpile12}
as
\begin{align}
\nonumber
	\ddot u_0({n},t) &= \alpha\delta^{\alpha-1}[u_0({n-1},t)-2u_0({n},t)+u_0({n+1},t)]\\
\nonumber
&\quad+ \frac{\alpha(\alpha-1)}{2}\delta^{\alpha-2}[(u_0({n-1},t)-u_0({n},t))^2  - (u_0({n},t)-u_0({n+1},t))^2]\\
&\quad+\mathcal{O}((|u_0({n-1},t)-u_0({n},t)|/\delta)^3)\,.
\label{1:longwave}
\end{align}
This expansion has
only a single nonlinear term in the retained orders, so we refer to
the physical regime in which this expansion is valid as a ``weakly nonlinear regime''.

In the long-wavelength limit, in which the characteristic size $L$ of a wave is large in comparison to the particle radius $R$,
\textcolor{black}{we write
\begin{equation}\label{e:taylor_u0}
	u_0(n,t) = u(x,t)\,, \quad x=2Rn\,, \quad u_0(n\pm1)=\e^{\pm2R\frac{\partial}{\partial x}}{u_0(n)}\,.
\end{equation}}
\textcolor{black}{Substituting~\eqref{e:taylor_u0} into~\eqref{1:longwave} and neglecting terms of order $O(u\delta^{\alpha-1}(R/L)^5[(R/L)+(u/\delta)])$ yields
\begin{equation}
	u_{tt}=c_0^2u_{xx}+2c_0\gamma u_{xxxx}-\sigma u_xu_{xx}\,,
\end{equation}
where
\begin{equation}
	c_0 = 2R\sqrt{\alpha\delta^{\alpha-1}}\,, \quad \gamma=\frac{c_0R^2}{6}\,,\quad \sigma=2(\alpha-1)\frac{c_0^2R}{\delta}\,.
\end{equation}
Introducing the transformations $\tau=c_0^3t$ and $\xi=x-c_0t$ and neglecting the higher-order term $u_{\tau\tau}$ gives
the KdV equation}
\begin{equation}
\label{1:kdv}
    c_0^3w_{\tau}+\gamma w_{\xi\xi\xi}+\frac{\sigma}{2c_0}ww_{\xi}=0\,,
\end{equation}
{where $w = -u_{\xi}$.}

We use solutions of \eqref{1:kdv} to approximate the leading-order behavior of $u(x,t)$ and
{then} obtain the leading-order behavior of $v(x,t)$ using~\eqref{1:leading}.
We are particularly interested in solutions that are perturbations of a leading-order solitary wave, and we thus select \textcolor{black}{$w(\xi,\tau)$ in~\eqref{1:kdv} to be the KdV solitary-wave solution
\begin{equation}\label{e:kdv_soliton}
	w(\xi,\tau)=A\,\mathrm{sech}^2
	\bigg[
	\sqrt{\frac{\sigma A}{24c_0\gamma}}\bigg(\xi-\frac{\sigma}{6c_0^4}A\tau\bigg)
	\bigg]\,.
\end{equation}
Integrating~\eqref{e:kdv_soliton} and transforming $\xi$, $\tau$ into $n$, $t$ yields the leading-order behavior}
\begin{align}
	u_0(n,t)= -\frac{\delta\epsilon}{\alpha-1} \tanh\left(\epsilon(n-c_\epsilon t)\right) + \mathcal{O}(\epsilon^{3/2})\,,
	\label{1:u00}
\end{align}
where
\begin{equation}
	\epsilon=\frac{2R}{L}\,,  \quad  c_{\epsilon} = \delta^{(\alpha-1)/2} \sqrt{\alpha}+\frac{\sqrt{\alpha}}{6}\epsilon^2\delta^{(\alpha-1)/2}\,, \quad L=\sqrt{\frac{24c_0\gamma}{\sigma A}}.
\end{equation}
The long-wavelength limit corresponds to $\epsilon \rightarrow 0$, and we thus refer to $\epsilon$ as the ``long-wavelength parameter''.

Motivated by \eqref{1:u00}, we use the co-moving frame $\xi = n - c_{\epsilon} t$. In terms of $\xi$, the leading-order behavior is
\begin{align}\label{1:u0}
	u_0(\xi) &= -\frac{\delta\epsilon}{\alpha-1} \tanh\left(\epsilon \xi\right) + \mathcal{O}(\epsilon^{3/2})\quad \mathrm{as} \quad \epsilon \rightarrow 0\,,\\
	v_0(\xi) &= u_0(\xi)\,.
	\label{1:v0}
\end{align}

The leading-order behavior has singularities at $\xi_{N\pm} = \pm\i(2N+1)\pi/2\epsilon$ for $N \in \mathbb{Z}$. The singularities that are closest to the real axis occur at $\xi_{0\pm} = \pm\i\pi/2\epsilon$ and are associated with singulants with the smallest values of $|\chi|$ when evaluated at real $\xi$. Consequently, these terms dominate the late-order behavior as $n \rightarrow \infty$. Near these singularities, we calculate that
\begin{align}\label{1:u0loc}
	v_0(\xi) \sim -\frac{\delta(\alpha-1)^{-1}  }{\xi - \xi_{0\pm}} \quad \mathrm{as} \quad \xi \rightarrow \xi_{0\pm}
\end{align}
and that $u_0$ has the same behavior as $v_0$ in the neighborhood of the singularity.

%%%%

\subsection{Late-Order Terms}

Writing the governing equations \eqref{1:gov1}--\eqref{1:gov2} in terms of $\xi$ and matching at each order of $\eta$ gives
\begin{align}\label{2:govser1}
	\nonumber c_{\epsilon}^2 u_j''(\xi) &= \alpha[u_j(\xi-1) - u_j(\xi))(\delta + u_0(\xi-1) - u_0(\xi)]^{\alpha-1} \\&\quad- \alpha[u_j(\xi) - u_j(\xi+1)][\delta + u_0(\xi) - u_0(\xi+1)]^{\alpha-1} - k [u_j(\xi) - v_j(\xi)] + \ldots\,,\\
	c_{\epsilon}^2 v_{j-1}''(\xi) &= k [u_j(\xi) - v_j(\xi)]\,,
	\label{2:govser2}
\end{align}
where we omit the terms that are products that include $u_{j-k}$ with $k>1$. These terms are subdominant with respect to the terms that we retain in the limit $j\to\infty$. We retain all terms that include $u_j$, $v_j$, and derivatives of $u_j$ and $v_{j-1}$. Given the general form of the factorial-over-power ansatz \eqref{e:lateorder_intro}, we conclude that the omitted terms do not contribute to the behavior of the asymptotic solution in our subsequent analysis. We will confirm this claim explicitly once we obtain the form of the late-order ansatz~\eqref{2:ansatz}.

In principle, one can apply~\eqref{2:govser1} and~\eqref{2:govser2} recursively to obtain terms in the series~\eqref{1:asympseries}
up to
arbitrarily large values of $j$. This process is challenging technically because it requires solving both a differential--difference equation~\eqref{2:govser1} and an algebraic equation~\eqref{2:govser2} at each order. Fortunately, this is not necessary for our analysis.
Additionally, obtaining terms up to arbitrary order does not reveal the presence of oscillations in the far field, where $\xi\to-\infty$, because the oscillations are exponentially small in the singularly perturbed limit $\eta\to0$.
Instead, we obtain the asymptotic form of the late-order terms as part of the exponential asymptotic process that we use to calculate the behavior of these oscillations.

The late-order ansatz consists of a sum of terms of the forms
\begin{equation}\label{2:ansatz}
	u_j \sim \frac{U(\xi)\Gamma(2j + \beta_1)}{\chi(\xi)^{2j + \beta_1}}\,, \quad v_j \sim \frac{V(\xi)\Gamma(2j + \beta_2)}{\chi(\xi)^{2j + \beta_2}} \quad \mathrm{as} \quad j \rightarrow \infty\,.
\end{equation}
To ensure that late-order terms have singularities at the same locations as the leading-order solution, we set $\chi=0$ at a particular choice of $\xi = \xi_{0\pm}$. 
{The full late-order expression is then
the sum of the late-order contributions from each of the singularities.}

For sufficiently large $j$, the terms of the asymptotic series~\eqref{2:ansatz} diverge in a factorial-over-power fashion, which confirms that $u_j\gg u_{j-k}$ and $v_j\gg v_{j-k}$ as $j\to\infty$ for $k > 0$. Additionally, inserting the late-order ansatz~\eqref{2:ansatz} into~\eqref{2:govser1} shows that only $\beta_1 + 2 = \beta_2$ produces a nontrivial asymptotic balance, implying that $u_j = \mathcal{O}(v_{j-1})$ as $j \rightarrow \infty$ and hence that $v_j \gg u_j$ as $j\to\infty$.

 Inserting the late-order ansatz \eqref{2:ansatz} into~\eqref{2:govser2} gives
 \begin{align}\nonumber
	 \frac{c_{\epsilon}^2 (\chi'(\xi))^2 V(\xi)\Gamma(2j + \beta_2)}{\chi(\xi)^{2j + \beta_2}}& - \frac{2c_{\epsilon}^2 \chi'(\xi) V'(\xi)\Gamma(2j + \beta_2-1)}{\chi(\xi)^{2j + \beta_2-1}}\\&
 	- \frac{c_{\epsilon}^2 \chi''(\xi) V(\xi) \Gamma(2j + \beta_2 - 1)}{\chi(\xi)^{2j + \beta_2 - 1}} + \cdots = -\frac{k V(\xi)\Gamma(2j + \beta_2)}{\chi(\xi)^{2j + \beta_2}} + \cdots\,,
 \end{align}
 where the omitted terms are no larger than $\mathcal{O}(v_{j-1})$ in the $j \rightarrow \infty$ limit.

Matching terms at $\mathcal{O}(v_j)$ in the $j \rightarrow \infty$
limit gives the singulant equation $c_{\epsilon}^2(\chi'(\xi))^2 = -k$. This implies that $\chi'(\xi) = \pm \i\sqrt{k}/c_{\epsilon}$, which we integrate to obtain
\begin{equation}\label{2:singulant}
	\chi(\xi) = \pm\frac{ \i \sqrt{k}(\xi-\xi_{0\pm})}{c_{\epsilon}}\,.
\end{equation}
Stokes switching can occur only if $\mathrm{Re}(\chi) > 0$, which corresponds to the positive sign choice for $\xi_{0+}$ and the negative sign choice for $\xi_{0-}$. Consequently, we retain these solutions and ignore the solutions that are associated with the remaining sign choices, as those can never appear in the asymptotic solution.

Matching terms at $\mathcal{O}(v'_{j-1})$ gives the prefactor equation $2 V'(\xi)\chi'(\xi) = 0$, so $V$ is a constant, with a value that depends on the choice of singularity. For clarity, we subsequently use $\Lambda_{\pm}$ to denote the constant prefactor that is associated with a singularity at $\xi = \xi_{0\pm}$. For the singular late-order behavior to be consistent with the local behavior of the leading-order solution \eqref{1:u0loc}, we calculate that $\beta_2 = 1$. In Appendix \ref{app:localw}, we perform a local expansion of the solutions $u(\xi)$ and $v(\xi)$ in a neighborhood of size $\mathcal{O}(\eps)$ near the singularity, and we use asymptotic matching to obtain
\begin{equation}\label{1:Lambda}
	\Lambda_+ = -\frac{\i\delta  \sqrt{k}}{(\alpha - 1) c_{\epsilon}}\,.
\end{equation}
We thereby fully determine the asymptotic behavior of $v_j$ in the $j \rightarrow \infty$ limit.

In Section~\ref{s:woodstokes}, we give a detailed calculation of the exponentially small oscillation that is associated with the singularity at $\xi=\xi_{0+}$.
At the conclusion of our analysis in Section~\ref{s:woodstokes}, we state the corresponding contribution that arises from the singularity at $\xi=\xi_{0-}$; this contribution is the complex conjugate of the contribution from $\xi=\xi_{0+}$.

%%%%%%

\subsection{Stokes Switching}\label{s:woodstokes}

We truncate the asymptotic series after $N$ terms to obtain
\begin{equation}\label{3:series}
	u(\xi) = \sum_{j=0}^{N-1} \eta^{2j} u_j(\xi) + S_N(\xi)\,,  \quad  v(\xi) = \sum_{j=0}^{N-1} \eta^{2j} v_j(\xi) + R_N(\xi)\,,
\end{equation}
where $S_N$ and $R_N$ are the remainder terms that we obtain by truncating the series. These remainder terms are exponentially small if we optimally truncate the series; we again denote the optimal truncation point as $N = N_{\mathrm{opt}}$.

\textcolor{black}{To optimally truncate the power series~\eqref{1:asympseries}, we follow the heuristic approach of~\cite{Boyd1999}. This approach requires truncating the series at the smallest term of the series. See~\cite{Boyd1999} for a detailed
discussion of the validity of this approach.
We locate the smallest term by taking the derivative of the
term $\eta^{2N}v_N(\xi)$ with respect to $N$ to obtain
\begin{equation} \label{e:optimal_derivative}
	\frac{\partial}{\partial N}\left|\eta^{2N}v_N\right|
\sim2\eta^{2N}\frac{|V|\Gamma(2N+1)}{|\chi|^{2N+1}}
\left(\log\eta-\log|\chi(\xi)|+\log(2N+1)\right)\,.
\end{equation}
The
smallest term occurs at the point at which the derivative term in~\eqref{e:optimal_derivative} is equal to $0$. This requires that
\begin{equation}
	\log(2N+1)\sim\log|\chi|-\log\eta\,.
\end{equation}
We thus obtain
\begin{equation} \label{e:optimal_result}
	N_{\mathrm{opt}}=|\chi|/2\eta+\omega\,,
\end{equation}
where we choose $\omega\in[0,1)$ such that $N_{\mathrm{opt}}$ is an integer. 
}

Inserting \eqref{3:series} into the governing equations \eqref{1:gov1}--\eqref{1:gov2} yields
\begin{align}
	c_{\epsilon}^2 S''(\xi) &\sim - k R_N(\xi)\,,\label{3:RSeq1}\\
	\eta^2 c_{\epsilon}^2 R''(\xi) + \eta^{2N} c_{\epsilon}^2 v_{N-1}''(\xi) & \sim -k R_N(\xi) \quad \mathrm{as} \quad \eta \rightarrow 0\,,
	\label{3:RSeq2}
\end{align}
where the omitted terms are smaller than those that we retain in the limit $\eta \rightarrow 0$.
Equation \eqref{3:RSeq2} decouples from \eqref{3:RSeq1},
so we can study it independently. Applying the late-order ansatz and rearranging gives
\begin{equation}\label{3:Seq}
	\eta^2 c_{\epsilon}^2 R_N'' + k R_N \sim -\frac{\Lambda\eta^{2N} (\chi')^2 \Gamma(2N + 1)}{\chi^{2N + 1}}\quad \mathrm{as} \quad \eta \rightarrow 0\,.
\end{equation}
The right-hand side of \eqref{3:Seq} is exponentially small except in a neighborhood around the Stokes curve. Away from the Stokes curve, we use the WKB method to obtain
\begin{align}
	R_N\sim C\e^{-\chi/\eta}\quad \mathrm{as} \quad \eta \rightarrow 0\,,
\label{3:WKB1}
\end{align}
where $C$ is a constant that we {need to determine}.

To capture the variation in the neighborhood of a Stokes curve, we adapt the form of~\eqref{3:WKB1} to include a Stokes-switching parameter $A(\xi)$, which is constant except in the neighborhood of the Stokes curve.
In such a neighborhood, $R_N$ takes the form
\begin{equation}\label{3:WKB}
	R_N(\xi) \sim A(\xi)\mathrm{e}^{-\chi/\eta}\quad \mathrm{as} \quad \eta \rightarrow 0\,.
\end{equation}
Inserting \eqref{3:WKB} into \eqref{3:Seq} and rearranging yields
 \begin{equation}
	\diff{A}{\xi} \sim \frac{ \Lambda\chi' \eta^{2N-1} \Gamma(2N+1)}{2\chi^{2N + 1}}\,\mathrm{e}^{\chi/\eta}\quad \mathrm{as} \quad \eta \rightarrow 0\,.
\end{equation}
Writing $N_{\mathrm{opt}}$ in terms of $\chi$, expanding the gamma function using Stirling's formula, and transforming to make $\chi$ the independent variable gives
\begin{equation}\label{this1}
	\diff{A}{\chi} \sim \frac{\Lambda\sqrt{\pi} \eta^{|\chi|/\eta + 2\omega-1}(|\chi|/\eta)^{|\chi|/\eta + 2\omega-1/2} \e^{-|\chi|/\eta}}{\sqrt{2} \chi^{|\chi|/\eta+2\omega+1}}\,\mathrm{e}^{\chi/\eta}\quad \mathrm{as} \quad \eta \rightarrow 0\,.
\end{equation}

We transform \eqref{this1} into polar coordinates using $\chi = \rho \mathrm{e}^{\i \theta}$ and consider variations in the angular direction. 
After some simplification, we obtain
\begin{equation}
	\diff{A}{\theta} \sim{\i\Lambda}\sqrt{\frac{\pi \rho}{2\eta^3}}\exp\left(\frac{\rho}{\eta}(\mathrm{e}^{\i\theta} - 1) - \frac{\i\theta \rho}{\eta} - 2 \i \omega \theta \right)\quad \mathrm{as} \quad \eta \rightarrow 0\,.
	\label{1:A}
\end{equation}
The right-hand side of~\eqref{1:A} is exponentially small in $\eta$, except in the neighborhood of $\theta=0$. Defining an inner region $\theta=\eta^{1/2}\bar{\theta}$, we find that
\begin{align}\label{this2}
	\frac{\d A}{\d\bar{\theta}}\sim\frac{\i\Lambda}{\eta }\sqrt{\frac{\pi \rho}{2}}\e^{-\rho\bar{\theta}^2/2}\,.
\end{align}
By integrating \eqref{this2}, we see that the behavior of $A$ as the Stokes curve is crossed is
\begin{align}\label{this3}
	A\sim\frac{\i\Lambda}{\eta}\sqrt{\frac{\pi }{2}}\int_{-\infty}^{\sqrt{\rho}\bar{\theta}}\e^{-s^2/2}\d s \quad \mathrm{as}\quad \eta\to0\,.
\end{align}
We evaluate the integral in \eqref{this3} and find that the difference between the values of $A$ on the two sides of the Stokes curve is
\begin{equation}
	[A]_-^+ \sim \frac{\i \pi \Lambda}{ \eta }\quad \mathrm{as} \quad \eta \rightarrow 0\,,
\end{equation}
where $[A]_-^+$ denotes the change in $A$ as the Stokes curve is crossed from $\theta < 0$ to $\theta > 0$.
Recalling~\eqref{1:Lambda} and~\eqref{3:WKB}, we find that the exponentially small contribution from $\xi=\xi_{0+}$ is
\begin{equation}\label{this4_before}
	[R_N]_-^+ \sim \frac{ \delta   \pi\sqrt{k}}{(\alpha-1) c_{\epsilon} \eta} \mathrm{e}^{-\i \sqrt{k}(\xi - \i \pi/2\epsilon)/c_{\epsilon}\eta} \quad \mathrm{as} \quad \eta \rightarrow 0\,.
\end{equation}
The exponentially small contribution from $\xi=\xi_{0-}$ is given by the complex conjugate of~\eqref{this4_before}.
Therefore, the total exponentially small contribution is
\begin{equation}\label{this4}
	[R_N]_-^+ \sim \frac{ \delta   \pi\sqrt{k}}{ (\alpha-1)c_{\epsilon} \eta} \mathrm{e}^{-\i \sqrt{k}(\xi - \i \pi/2\epsilon)/c_{\epsilon}\eta} + \mathrm{c. c.} \quad \mathrm{as} \quad \eta \rightarrow 0\,,
\end{equation}
where c.c. denotes the complex conjugate. We express \eqref{this4} in terms of trigonometric functions and thus write
\begin{equation}
	[R_N]_-^+ \sim\frac{2\delta   \pi\sqrt{k}}{(\alpha-1)\eta c_{\epsilon}}
\exp\left(-\frac{\pi\sqrt{k}}{2 c_{\epsilon}\epsilon\eta}\right)
\cos\left(\frac{ \xi\sqrt{k}}{c_{\epsilon}\eta}\right)\quad \mathrm{as} \quad \eta \rightarrow 0\,.
\label{e:woodpile_sn}
\end{equation}
We are considering one-sided nanoptera for which $R_N = 0$
in the undisturbed upstream region. This corresponds to requiring that $R_N = 0$ for $\theta < 0$ and thus that $R_N \sim [R_N]_-^+$ for $\theta > 0$.
If we were studying symmetric two-sided nanoptera, we would instead set $R_N \sim -\tfrac{1}{2}[R_N]_-^+$ for $\theta < 0$ and $R_N \sim \tfrac{1}{2}[R_N]_-^+$ {for $\theta > 0$}. This is consistent with the jump
in \eqref{e:woodpile_sn}, and it yields an oscillation amplitude that is half of that of
the corresponding one-sided nanopteron.

%%%%%

\subsection{Comparison of our Asymptotic and Computational Results}\label{s:woodnum}

We compare the amplitude of the oscillations that we predict using our asymptotic analysis to that from numerical simulations. We employ a symplectic integrator in the form of a velocity Verlet algorithm~\cite{Verlet,Allen}, which is convenient for studying chains of particles.
This approach is designed to conserve the energy of a system, which is not true of many common numerical methods, such as Runge--Kutta algorithms. The velocity Verlet algorithm uses the discretization
\begin{equation}
	\vec{x}_{n+1}=\vec{x}_n+\vec{v}_n\Delta t+\frac{1}{2} \vec{a}_{n}(\Delta t)^2\,,  \quad  \vec{v}_{n+1}=\vec{v}_n+\frac{1}{2}(\vec{a}_n+\vec{a}_{n+1})\Delta t\,,
\end{equation}
where $\Delta t$ is the size of the time step
{and $\vec{x}_n$, $\vec{v}_n$, and $\vec{a}_n$ are vector quantities that
encode
the displacements, velocities, and accelerations of the particles at time $t=t_n$}. In a woodpile chain, we obtain the acceleration vector $\vec{a}_n$ as an algebraic function of the displacement vector $\vec{x}_n$ using \eqref{1:gov1}--\eqref{1:gov2}.

We truncate the domain and impose periodic boundary conditions. Instead of directly computing the absolute displacements and velocities of the particles, we perform our computations using the relative displacements and velocities, which we compute by calculating the differences in the positions and velocities of adjacent particles. This is convenient for our computations because the far-field behavior of the leading-order traveling wave approaches $0$ in both directions in these coordinates. The relative displacements of the heavy particles and light particles are $r_1$ and $r_2$, respectively, where
\begin{equation}
	r_1(n,t) = u(n+1,t) - u(n,t)\,, \quad  r_2(n,t) = v(n+1,t) - v(n,t)\,.
\end{equation}
In relative coordinates, the governing equations are
\begin{align}
	\ddot{r_1}(n,t) &= 2 [\delta  - r_1(n,t)]_+^\alpha
-[\delta  - r_1(n-1,t) ]_+^\alpha -[\delta  - r_1(n+1,t)]_+^\alpha  - k [r_1(n,t) - r_2(n,t)]\,,\\
	\eta^2 \ddot{r_2}(n,t) &= k [r_1(n,t) - r_2(n,t)]\,.
\end{align}

The truncated domain has $M=2^{10}$ particles, whose indices are $n \in \{ -M/2 + 1, \ldots, M/2\}$. The initial condition is the leading-order traveling-wave solution~\eqref{1:u0}--\eqref{1:v0}. To avoid interactions between the far-field oscillations and the leading-order traveling wave in the periodic domain, we apply a window function to the solution, as in~\cite{Giardetti,Lustri1}, at each iteration. We obtain the window by multiplying $r(n,t)$ and $\dot{r}(n,t)$ by a function $W(n-n_{\max}+M/8)$, where
\begin{equation}\label{e:window}
W(k) = \left\{
        \begin{array}{ll}
            1\,, & \quad |K| \leq \tfrac{5M}{16} \\
            1 - \tfrac{8}{M}\left(|K| - \tfrac{5M}{16}\right)\,, & \quad \tfrac{5M}{16} < |K| \leq \tfrac{7M}{16} \\
            0\,, & \quad\tfrac{7M}{16} < |K| \leq \tfrac{M}{2}
        \end{array}
    \right.
\end{equation}
and $n_{\max}$ denotes the location of the maximum of the leading-order solution.

In~\cite{Giardetti}, it was argued that a window of this form cannot affect the behavior of the main wave or trailing oscillations in simulations of FPUT systems, as any disturbances that are associated with the windowing must travel more slowly than the leading-order solitary wave and hence than the window itself (see \cite{FrieseckePego}). We do not perform a comparable analysis for a woodpile chain, but we do not observe any discernible differences in the amplitudes of the trailing oscillations as a result of our windowing.

\begin{figure}
\centering
\includegraphics[width=0.5\textwidth]{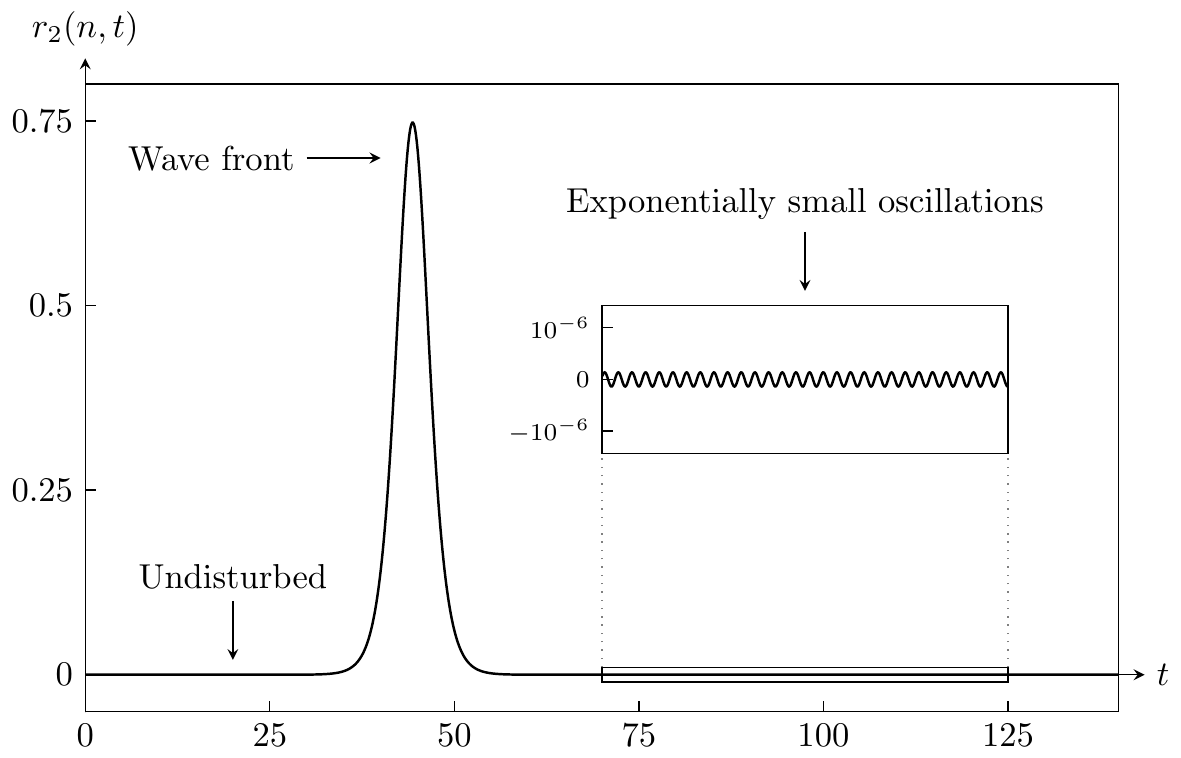}

\caption{Numerically-calculated relative displacement of a single light particle at some index $n$ in a woodpile chain with an interaction exponent of $\alpha=1.5$. The mass-ratio parameter is $\eta = 0.4$, the precompression
of the chain {gives} $\delta = 5$, and the long-wavelength parameter is $\eps = 0.3$.
}
   \label{f:numerics}
\end{figure}

\begin{figure}%[tb]
\centering

\subfloat{
\includegraphics{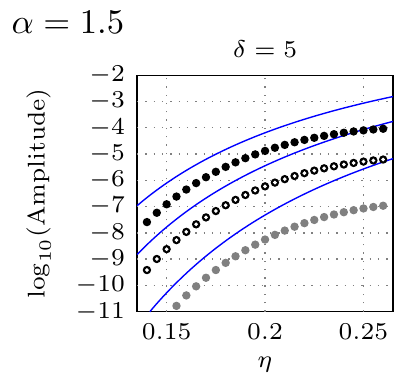}
}
\subfloat{
\includegraphics{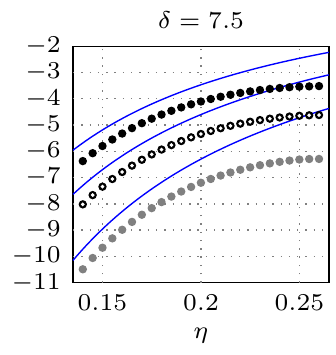}
}
\subfloat{
\includegraphics{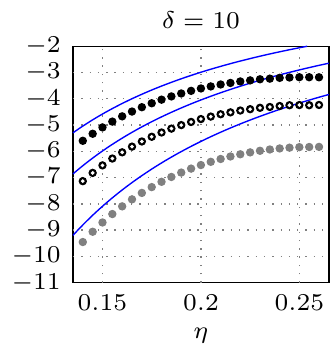}
}
\subfloat{
\includegraphics{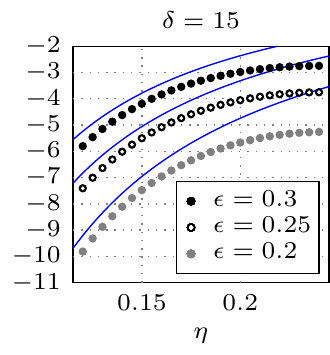}
}

\subfloat{
\includegraphics{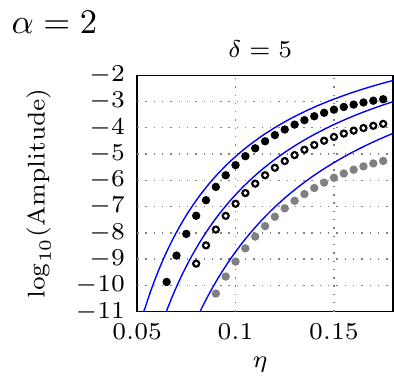}
}
\subfloat{
\includegraphics{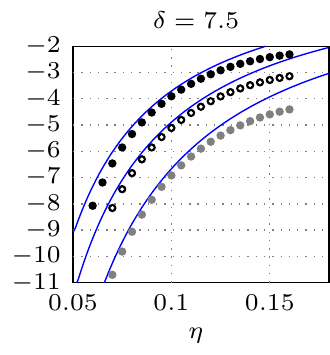}
}
\subfloat{
\includegraphics{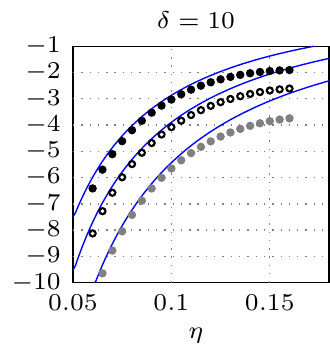}
}
\subfloat{
\includegraphics{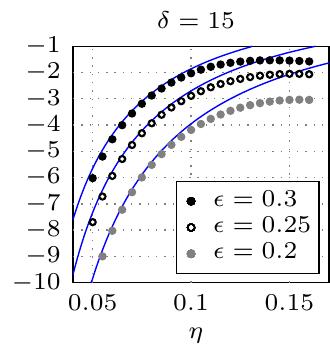}
}

\subfloat{
\includegraphics{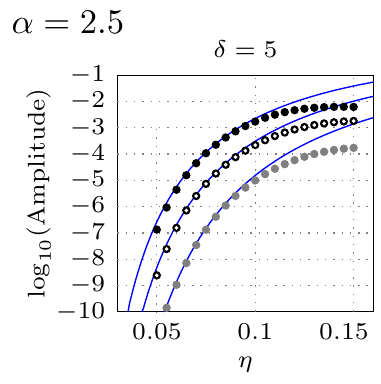}
}
\subfloat{
\includegraphics{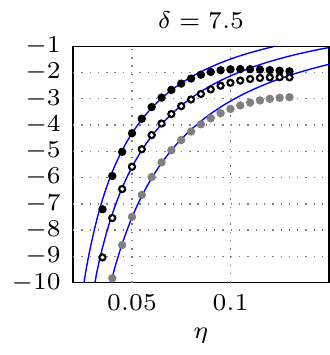}
}
\subfloat{
\includegraphics{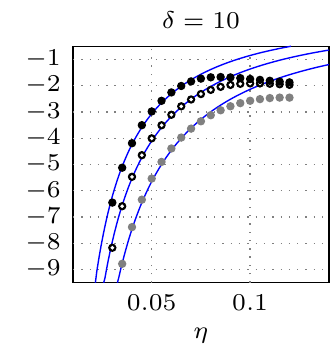}
}
\subfloat{
\includegraphics{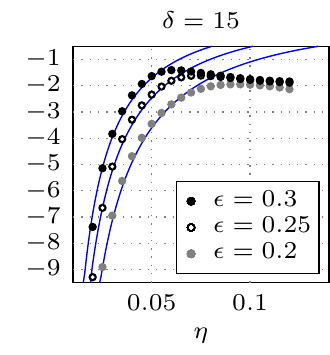}
}

\caption{Comparison of our asymptotic approximations and numerical simulations of the amplitude of exponentially small oscillations in the far field (i.e., as $\xi\to-\infty$) for woodpile chains with interaction exponents of (a) $\alpha = 1.5$, (b) $\alpha = 2$, and (c) $\alpha = 2.5$. For each $\alpha$, we show simulations for precompressions of $\delta=5$, $\delta=7.5$, $\delta=10$, and $\delta=15$. For each choice of $\delta$, we set the long-wavelength parameter to be $\epsilon = 0.2$, $\epsilon = 0.25$, and $\epsilon = 0.3$. The accuracy of our asymptotic approximation improves for progressively smaller values of the mass-ratio parameter $\eta$, as expected for an $\eta \rightarrow 0$ asymptotic approximation. Additionally, the approximation improves as we increase $\delta$. This is a consequence of using a leading-order approximation for the weakly nonlinear regime; decreasing $\delta$ approaches a configuration that needs to be treated as strongly nonlinear. Our approximation also improves as we increase $\alpha$, although the reason for this improvement is not apparent.
}\label{F:Woodpile}
\end{figure}

{In Fig.~\ref{f:numerics}, we show a numerically-calculated profile of one particle in a woodpile chain. In this simulation, we use an interaction exponent of $\alpha=1.5$, a mass-ratio parameter of $\eta = 0.4$, a precompression of $\delta = 5$, and a long-wavelength parameter of $\eps = 0.3$. The curve in the figure shows the relative displacement of a single light particle
over a range of times $t$. The particle is initially at rest before the leading-order solitary wave reaches the particle, causing it to become displaced. Once this solitary wave has passed the particle, the particle continues to oscillate with an exponentially small amplitude. In Fig.~\ref{F:Woodpile}, we measure this oscillation amplitude from numerical simulations and compare it to our asymptotic approximations for a range of
parameter choices.}

In our numerical simulations, we calculate the amplitude of the far-field waves for interaction exponents of $\alpha=1.5$, $\alpha=2$ and $\alpha=2.5$. For each $\alpha$, we show simulations with precompressions of $\delta = 5$, $\delta = 7.5$, $\delta = 10$, and $\delta =15$. For each choice of $\delta$, we set the long-wavelength parameters to be $\eps = 0.2$, $\eps = 0.25$, and $\eps = 0.3$. We choose the values of $\eta$ so that the far-field waves are large enough to be detected numerically, but small enough that our small-$\eta$ asymptotic approximation
accurately describes the behavior of the solution. We run each of our simulations for enough time
so that any time variation in the far-field oscillation amplitude is not visible in the solution. We show the results of these simulations, which we compare to the asymptotic results from \eqref{e:woodpile_sn}, in Fig.~\ref{F:Woodpile}.

It is apparent from Fig.~\ref{F:Woodpile} that our asymptotic approximation captures the far-field wave amplitude effectively for a wide range of the parameters and that the approximation error decreases as $\eta \rightarrow 0$, as we expect from the asymptotic nature of the approximation. For a fixed $\epsilon$, our asymptotic approximation becomes progressively more accurate for progressively larger $\delta$. However, for progressively smaller $\delta$, the system approaches the strongly nonlinear regime and the long-wavelength approximation becomes less accurate, which in turn causes the late-order approximation to become less accurate. Our asymptotic approximation also appears to become more accurate for progressively larger $\alpha$, but the reason for this improvement is not apparent from our analysis.

By comparing our numerical computations and our asymptotic results, we conclude that we can explain the oscillations that appear in woodpile chains as a consequence of Stokes phenomena
and that the employed exponential asymptotic analysis
is a useful technique for studying nanoptera in these chains.

%%%%%

\section{A Singularly Perturbed Diatomic Hertzian Chain}
\label{s:diatomic}

We now consider a singularly perturbed diatomic Hertzian chain. This system is related to the woodpile chains that we analyzed in Section \ref{s:woodpile}. Once we derive a leading-order solution for a traveling wave in a diatomic Hertzian chain, we can perform a similar exponential asymptotic analysis. Obtaining the singulant equation for diatomic Hertzian chains requires solving a complicated integral expression; this is significantly more
complicated
than the corresponding step in our analysis of woodpile chains.

We consider a diatomic Hertzian chain, which is governed by~\eqref{e:diatomic01}--\eqref{e:diatomic02}. We denote the masses in the chain by $m(j)=m_1$ for even $j$ and $m(j)=m_2$ for odd $j$.
We consider a small ratio between the masses of the two types of particles. This choice amounts to a precompressed Hertzian analog of the diatomic Toda and FPUT chains that were studied in \cite{Lustri, Lustri1}. Those previous studies used far-field asymptotic behavior to identify an orthogonality condition, which yields
values of the mass-ratio parameter $\eta$ that cause the far-field waves to cancel and produce a genuine solitary wave.

We derive an asymptotic approximation in a diatomic Hertzian chain to identify an orthogonality condition that finds
some (but not all) values of $\eta$ that produce solitary waves. Specifically, our asymptotic approach systematically fails to identify every second value of $\eta$ that causes the far-field oscillations to cancel.

Much of our analysis in this section is similar to that in Section \ref{s:woodpile} and in \cite{Lustri, Lustri1}. Therefore, we present only an outline of our calculations.

The governing equations for a diatomic chain of particles is
\begin{align}
	m_1 \ddot u(2n,t) &= [\delta_0+ v(2n-1,t) - u(2n,t)]_+^{\alpha} - [\delta_0+u(2n,t) - v(2n+1,t)]_+^{\alpha}\,,\label{e:diatomic01}\\
	m_2 \ddot v(2n+1,t) &= [\delta_0+ u(2n,t) - v(2n+1,t)]_+^{\alpha} - [\delta_0+v(2n+1,t) - u(2n+2,t)]_+^{\alpha}\,,\label{e:diatomic02}
\end{align}
where $u$ and $v$, respectively, represent the displacement of even and odd particles, the $+$ subscript has the same meaning as in \eqref{1:gov1}--\eqref{1:gov2}, and we set $\alpha = 3/2$ so that we consider Hertzian interactions.\footnote{
For diatomic Hertzian chains, the form of the singulant depends
on the value of $\alpha$. By contrast, for woodpile chains, the singulant is independent of $\alpha$. Therefore, in Section \ref{s:woodpile}, we permitted $\alpha$ to take any value that produces leading-order solitary waves, as it did not change our subsequent analysis. 
For diatomic chains, we must pick a specific value of $\alpha$.
Therefore, we restrict our attention to Hertzian interactions, which are the most common case realized in existing laboratory experiments.}
We apply the same scalings as we did for woodpile chains, so we let $u = m_1^2 \hat{u}$ and $v= m_1^2 \hat{v}$. We rewrite \eqref{e:diatomic01}--\eqref{e:diatomic02} {in terms of our scaled variables and a scaled precompression parameter $\hat{\delta} = \delta_0/m_1^2$}. Setting $\eta^2 = m_2/m_1$ yields the scaled system
\begin{align}
	 \ddot{\hat{u}}(2n,t) &= [\hat{\delta}+ \hat{v}(2n-1,t) - \hat{u}(2n,t)]_+^{3/2} - [\hat{\delta}+\hat{u}(2n,t) - \hat{v}(2n+1,t)]_+^{3/2}\,,\label{e:diatomic1}\\
	\eta^2 \ddot{\hat{v}}(2n+1,t) &= [\hat{\delta}+ \hat{u}(2n,t) - \hat{v}(2n+1,t)]_+^{3/2} - [\hat{\delta}+\hat{v}(2n+1,t) - \hat{u}(2n+2,t)]_+^{3/2}\,.\label{e:diatomic2}
\end{align}
{As in the woodpile chains, we assume that the precompression is sufficiently strong that the particles remain in contact and thus that the quantities in the square brackets in~\eqref{e:diatomic1}--\eqref{e:diatomic2} are never negative.}
Therefore, we omit the $+$ subscript in our subsequent notation.
In our analysis of the scaled system \eqref{e:diatomic1}--\eqref{e:diatomic2}, we also omit the hats from the variables and the parameters for notational convenience.

As in the woodpile chains of Section \ref{s:woodpile}, we study traveling-wave solutions of \eqref{e:diatomic1}--\eqref{e:diatomic2} when the mass-ratio parameter $\eta$ is small. Specifically, we consider $0 < \eta \ll 1$. As before, the solution consists of a localized wave core 
and exponentially small, non-decaying oscillations. We again use exponential asymptotic methods to study these oscillations.

%%%%

\subsection{Leading-Order Solution}
\label{s:leading}

The first step to construct a nanopteron solution of \eqref{e:diatomic1}--\eqref{e:diatomic2} is to find a leading-order solitary wave. Therefore, we begin by expanding the solution $u(x,t)$ and $v(x,t)$ using an asymptotic power series in $\eta^2$ in the limit $\eta \rightarrow 0$. This yields
\begin{align}
	u(n,t)\sim\sum_{j=0}^\infty \eta^{2j}u_j(n,t) \,,  \quad v(n,t)\sim\sum_{j=0}^\infty \eta^{2j}v_j(n,t)\,.
\label{e:asympseries}
\end{align}

Inserting the series expansions \eqref{e:asympseries} into~\eqref{e:diatomic2} and matching at leading order in the $\eta \rightarrow 0$ limit gives
\begin{align}
	v_0(2n+1,t) =\tfrac{1}{2}[u_0(2n,t)+u_0(2n+2,t)]\,.
\label{e:diatomic22}
\end{align}
Inserting~\eqref{e:diatomic22} into~\eqref{e:diatomic1} yields
\begin{align}
	\ddot u_0(2n,t)=\bigg(\delta+\frac{u_0({2n-2},t)-u_0({2n},t)}{2}\bigg)^{3/2}-\bigg(\delta +\frac{u_0({2n},t)-u_0({2n+2},t)}{2}\bigg)^{3/2}\,.
\label{e:diatomic12}
\end{align}

Using a similar analysis as in Section \ref{s:woodLO}, we obtain an approximation to the leading-order solution by assuming that the displacement of a particle is small in comparison to the precompression parameter $\delta$ and that the characteristic length scale of the leading-order wave is large in comparison to the radius of the spherical particles. We again define a long-wavelength parameter $\epsilon$.
The leading-order solution is approximately
\begin{align}
	u_0(\xi)= -4\delta  \epsilon \tanh\left(\epsilon\xi/2\right)+\mathcal{O}(\epsilon^{3/2})\,,  \quad  v_0(\xi)=\tfrac{1}{2}(u_0(\xi+1)+u_0(\xi-1))\,,
\label{e:leading}
\end{align}
where the variable $\xi$ defines the co-moving frame
\begin{align}
	\xi = n-c_\epsilon t\,,  \quad  c_{\epsilon} = \delta^{1/4} \sqrt{3}+\frac{\delta^{1/4}\epsilon^2}{2\sqrt{3}}\,.
\end{align}

\begin{figure}
\centering
\includegraphics{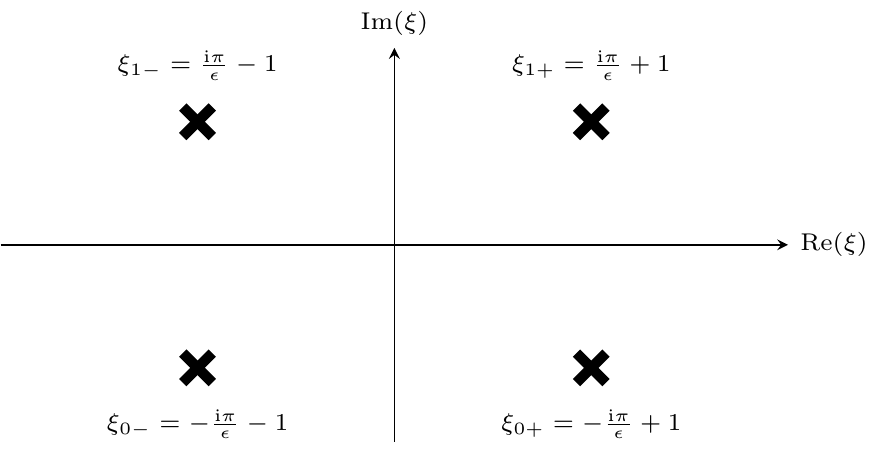}

\caption{Singularities of $v_0(\xi)$ in \eqref{e:xi} that contribute to the asymptotic form of the far-field oscillations of~\eqref{e:diatomic_oscillation}.   We give a detailed discussion of the analysis of the oscillations that arise from the singularity at $\xi = \xi_{1-}$. We then state the contributions from the other three singularities.
}
   \label{f:singularity}
\end{figure}

We now extend the leading-order solution~\eqref{e:leading} into the complex plane. We see that $v_0(\xi)$ is singular at $\xi=\xi_{N\pm}$ in the complex plane, where
\begin{align}
	\xi_{N\pm}=\frac{{(2N-1)}\pi\i}{\epsilon}\pm1 \,, \quad N\in\mathbb{Z}\,.
\label{e:xi}
\end{align}
The behavior near the singularities is
\begin{align}
	v_0(\xi)\sim-4\delta(\xi-\xi_{N\pm})^{-1} \quad \mathrm{as} \quad \xi\to\xi_{N\pm}\,.
\label{e:v0sing}
\end{align}
As with woodpile chains,
the dominant contribution to the late-order behavior arises from the singularities that are closest to the real axis; this occurs when $N=0$ and $N=1$ in~\eqref{e:xi}. We illustrate the positions of these singularities in Fig.~\ref{f:singularity}. In our subsequent analysis, we show the details for calculating the contribution from the singularity at $\xi=\xi_{1-}$.
We then state the analogous results for the contributions from the singularities at $\xi=\xi_{1+}$, $\xi_{0-}$, and $\xi_{0+}$.

%%%%

\subsection{Terms in the Late-Order Series}
\label{s:late}

Inserting the series \eqref{e:asympseries} into the system \eqref{e:diatomic1}--\eqref{e:diatomic2} and matching at each order of $\eta$ gives a recurrence relation for $j\geq2$. The recurrence relation is
\begin{align}  \label{2:govern1}
	\nonumber c_{\epsilon}^2 u_j''(\xi) = &\tfrac{3}{2}(v_j(\xi-1) - u_j(\xi))(\delta+ v_0(\xi-1) - u_0(\xi))^{1/2} \\
\nonumber&- \tfrac{3}{2}(u_j(\xi) - v_j(\xi+1))(\delta+ u_0(\xi) - v_0(\xi+1))^{1/2} \\
\nonumber&+\tfrac{3}{4}(v_{j-1}(\xi-1) - u_{j-1}(\xi))(v_1(\xi-1) - u_1(\xi))[\delta+(v_0(\xi-1) - u_0(\xi))]^{-1/2}\\
&-\tfrac{3}{4}(u_{j-1}(\xi) - v_{j-1}(\xi+1))(u_1(\xi) - v_1(\xi+1))[\delta+(u_0(\xi) - v_0(\xi+1))]^{-1/2}+\cdots \,,\\
\nonumber c_{\epsilon}^2 v_{j-1}''(\xi) =& \tfrac{3}{2}(u_j(\xi-1) - v_j(\xi))(\delta+ u_0(\xi-1) - v_0(\xi))^{1/2} \\
\nonumber&- \tfrac{3}{2}(v_j(\xi)- u_j(\xi+1))(\delta+ v_0(\xi) - u_0(\xi+1))^{1/2} \\
\nonumber&+\tfrac{3}{4}(u_{j-1}(\xi-1) - v_{j-1}(\xi))(u_1(\xi-1) - v_1(\xi))[\delta+(u_0(\xi-1) - v_0(\xi))]^{-1/2}\\
	&-\tfrac{3}{4}(v_{j-1}(\xi) - u_{j-1}(\xi+1))(v_1(\xi) - u_1(\xi+1))[\delta+(v_0(\xi) - u_0(\xi+1))]^{-1/2}+\cdots\,,\label{2:govern2}
\end{align}
where we omit the terms that are products that include both $u_{j-k}$ and $v_{j-k}$ with $k>1$. The terms that we retain are those that contain $u_{j-1}$, $v_{j-1}$, $u_j$, and $v_j$. All of the omitted terms are subdominant in comparison to the retained terms as $j\to\infty$.

We again use a factorial-over-power ansatz to approximate the late-order terms, so we write
\begin{equation}
	u_j \sim \frac{U(\xi)\Gamma(2j + \beta_1)}{\chi(\xi)^{2j + \beta_1}}\,, \quad v_j \sim \frac{V(\xi)\Gamma(2j + \beta_2)}{\chi(\xi)^{2j + \beta_2}} \quad \mathrm{as} \quad j \rightarrow \infty\,,
\label{e:lateorder1}
\end{equation}
where $\beta_1$ and $\beta_2$ are constants. Inserting the late-order ansatz~\eqref{e:lateorder1} into~\eqref{2:govern1} shows that only $\beta_1 = \beta_2-2$ produces a nontrivial asymptotic balance.

%%%%%

\subsubsection{Calculating $\chi$}\label{s:chi}

\begin{figure}
\centering
\includegraphics{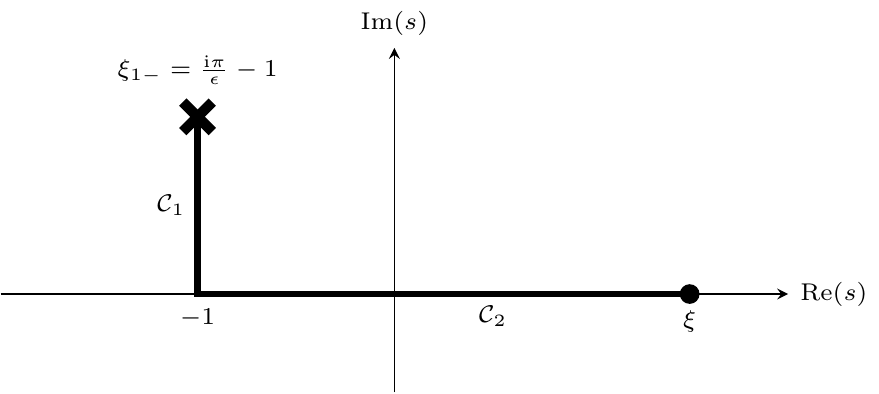}

\caption{The contour of integration for \eqref{e:chiint} that connects the singularity at $s = \xi_{1-}$ with the point $s = \xi$.
We indicate the location of the singularity using a cross, and we show the contour using a thick black curve. We divide the contour into a vertical component $\mathcal{C}_1$ and a horizontal component $\mathcal{C}_2$. The contribution to the contour integral from $\mathcal{C}_1$ has both real and imaginary components, whereas the contribution to the integral from $\mathcal{C}_2$ is purely imaginary. This implies that $\mathrm{Re}(\chi)$ is constant for real-valued $\xi$.}
\label{f:contour}
\end{figure}

Inserting the ansatz~\eqref{e:lateorder1} into the recurrence relation~\eqref{2:govern1}--\eqref{2:govern2} and matching at $\mathcal{O}(v_j)$ as $j\to\infty$ gives
\begin{equation}
	c_\epsilon^2(\chi')^2 = -3\bigg(\delta+\frac{u_0(\xi-1) - u_0(\xi+1)}{2}\bigg)^{1/2}\,.
\label{e:chisquare}
\end{equation}
Integrating~\eqref{e:chisquare} and recalling that $\chi=0$ at the singularity location $\xi=\xi_{1-}$, we obtain
\begin{equation}
	\chi=\pm\frac{\sqrt{3}\i}{c_\epsilon}\int_{\mathcal{C}}\bigg(\delta +\frac{(u_0(s-1) - u_0(s+1))}{2}\bigg)^{1/4}\d s\,,
\label{e:chiint}
\end{equation}
where {$\mathcal{C}$} is a contour from $\xi_{1-}$ to $\xi$.
Although one can select any contour, it is convenient to divide the contour into a vertical component $\mathcal{C}_1$ and a horizontal component $\mathcal{C}_2$. We show our contour in Fig.~\ref{f:contour}.

The contribution to the integral in \eqref{e:chiint} from $\mathcal{C}_1$ has both real and imaginary components, but the contribution to the integral from $\mathcal{C}_2$ is purely imaginary. The real and imaginary parts of $\chi$ are
\begin{align} \label{e:chiintrealimag}
	\mathrm{Re}(\chi) = \pm\mathrm{Re}\bigg\{\frac{\sqrt{3}\i}{c_\epsilon}\int_{\mathcal{C}_1}\bigg(\delta+&\frac{u_0(s-1) - u_0(s+1)}{2}\bigg)^{1/4}\d s\bigg\}\,,\\
\nonumber\mathrm{Im}(\chi) = \pm\mathrm{Im}\bigg\{\frac{\sqrt{3}\i}{c_\epsilon}\int_{\mathcal{C}_1}\bigg(\delta+&\frac{u_0(s-1) - u_0(s+1)}{2}\bigg)^{1/4}\d s\bigg\}\\
                    & \pm\frac{\sqrt{3}\i}{c_\epsilon}\int_{\mathcal{C}_2}\bigg(\delta+\frac{u_0(s-1) - u_0(s+1)}{2}\bigg)^{1/4}\d s\,,
\label{e:chiintrealimag1}
\end{align}
{where the sign choices are either all positive or all negative.} We see that $\mathrm{Re}(\chi)$ is constant for real-valued $\xi$, so the far-field oscillations have a constant amplitude. Because $|u_0(\xi-1)-u_0(\xi+1)| \rightarrow 0$ as $|\xi| \rightarrow \infty$, we also see that $\chi'(\xi)$ tends to a constant value and that $\mathrm{Im}(\chi)$ depends linearly on $\xi$ in this limit. Consequently, the associated far-field oscillations must tend to a constant wavelength.

Stokes switching can occur only if $\mathrm{Re}(\chi)>0$, which corresponds to the positive signs {for $\xi_{1-}$} in both~\eqref{e:chiintrealimag} and~\eqref{e:chiintrealimag1}. Therefore, we
restrict our analysis to this choice of signs.
For our analysis,
it is helpful to have an asymptotic expression for the local behavior of $\chi$ near the singular point $\xi_{1-}$.
From direct calculation using local expressions for $u_0(\xi+1)$ and $u_0(\xi-1)$, which we give explicitly in~\eqref{e:UVinner}, we obtain
\begin{align}
	c_\epsilon^2(\chi')^2\sim-6\sqrt{\delta}(\xi-\xi_{1-})^{-1/2} \quad \mathrm{as} \quad \xi\to\xi_{1-}\,.
\label{e:chiloc}
\end{align}
We rearrange this local expression to give
\begin{align}
	 \chi'\sim\pm\frac{\sqrt{6}\delta^{1/4}\i}{c_\epsilon}(\xi-\xi_{1-})^{-1/4} \quad \mathrm{as} \quad \xi\to\xi_{1-}\,.
\label{e:chiloc1}
\end{align}
We choose the positive sign to be consistent with the full expression for $\chi$, and we integrate to obtain
\begin{align}
	 \chi\sim\frac{4\sqrt{6}\delta^{1/4}\i}{3c_\epsilon}(\xi-\xi_{1-})^{3/4} \quad \mathrm{as} \quad \xi\to\xi_{1-}\,.
\label{e:chiloc2}
\end{align}

%%%%%%

\subsubsection{Calculating $V$ and $\beta_2$}

Matching~\eqref{2:govern1}--\eqref{2:govern2} at order $\mathcal{O}(v_{j-1}')$ as $j \rightarrow \infty$ gives
\begin{align}
	V=\frac{\Lambda_{1-}}{\sqrt{\chi'(\xi)}}\,,
\label{e:V}
\end{align}
where $\Lambda_{1-}$ is a constant that we can determine by taking an inner expansion of the solution in the neighborhood of $\xi=\xi_{1-}$ and matching the outer limit of this expansion with the inner limit of the late-order ansatz~\eqref{e:lateorder1}. We need to determine the value of $\beta_2$ in~\eqref{e:lateorder1} to perform this inner analysis.

To determine $\beta_2$, we combine~\eqref{e:chiloc1} and~\eqref{e:V} to obtain a local expression for $V$. This expression is
\begin{align}
	V\sim\frac{\Lambda_{1-} c_\epsilon^{1/2}}{6^{1/4}\delta^{1/8}\sqrt{\i}}(\xi-\xi_{1-})^{1/8} \quad \mathrm{as} \quad \xi\to\xi_{1-}\,.
\label{e:Vloc}
\end{align}
Substituting~\eqref{e:chiloc2} and~\eqref{e:Vloc} into~\eqref{e:lateorder1} yields a local expression for the late-order ansatz:
\begin{align}
	v_j(\xi)\sim\frac{\Lambda_{1-} c_\epsilon^{1/2}(\xi-\xi_{1-})^{1/8}\Gamma(2j+\beta_2)}
{6^{1/4}\delta^{1/8}\sqrt{\i}(4\sqrt{6}\delta^{1/4}\i(\xi-\xi_{1-})^{3/4}/3c_\epsilon)^{2j+\beta_2}}\,.
\label{e:v_outer}
\end{align}
The leading-order solution $v_0$ \eqref{e:leading} has a singularity of order $1$
at $\xi=\xi_{1-}$. For \eqref{e:v_outer} to be consistent with repeated differentiation of the leading-order behavior, we require that $3\beta_2/4-1/8=1$, yielding $\beta_2=3/2$.

Finally, we need to determine $\Lambda_{1-}$. The late-order expansion \eqref{e:lateorder1} breaks down in the neighborhood of $\xi = \xi_{1-}$. 
In this neighborhood, we determine a local expansion of $v_j(\xi)$.
By matching the inner limit of the late-order behavior [which is given in \eqref{e:v_outer}] with the outer limit of the local expansion, we
obtain $\Lambda_{1-}$. We show the computational portion of this procedure in Appendix \ref{app:local1} and obtain $c_{\epsilon}^2\Lambda_{1-}/\delta^{3/2} \approx 38.41$.

%%%%%

\subsection{Calculations of the Remainders}
\label{s:stokes}

As in Section~\ref{s:woodpile}, we truncate the asymptotic series after $N$ terms. This yields
\begin{align}
	u(\xi)=\sum_{j=0}^{N-1}\eta^{2j}u_j(\xi)+S_N(\xi)\,,  \quad  v(\xi)=\sum_{j=0}^{N-1}\eta^{2j}v_j(\xi)+R_N(\xi)\,,
\label{e:truncated}
\end{align}
where $S_N$ and $R_N$ are the remainder terms that we obtain by truncating the series. The optimal truncation point (which we find using the same method as in Section~\ref{s:woodstokes}) is $N_{\mathrm{opt}}=|\chi|/2\eta+\omega$, where we choose $\omega\in[0,1)$ to ensure that $N_{\mathrm{opt}}$ is an integer. Note that $N_{\mathrm{opt}}\to\infty$ as $\eta\to0$.

We insert the truncated series expressions from~\eqref{e:truncated} into the governing equation~\eqref{e:diatomic2}. Using the recursion relation~\eqref{2:govern2} and the late-order ansatz~\eqref{e:lateorder1}, we obtain
\begin{align}
	 c_\epsilon^2\eta^2R_N''(\xi)-c_\epsilon^2\chi'(\xi)^2R_N(\xi)\sim-\frac{\eta^{2N}\chi'(\xi)^2V(\xi)\Gamma(2N+3/2)}{\chi(\xi)^{2N+3/2}} \quad \mathrm{as} \quad \eta \rightarrow 0
\label{e:R_N1}
\end{align}
after some algebra.

The right-hand side of~\eqref{e:R_N1} is exponentially small and one can neglect it, except in the neighborhood of the curve $\mathrm{Im}(\chi) = 0$, which is the Stokes curve. 
To capture the behavior of the remainder $R_N$ near the Stokes curve,
we write it
using the same adapted WKB ansatz as in~\eqref{3:WKB}.
That is,
\begin{equation}\label{4:WKB}
	R_N(\xi) \sim A(\xi)\mathrm{e}^{-\chi/\eta}\quad \mathrm{as} \quad \eta \rightarrow 0\,,
\end{equation}
where $A(\xi)$ is a Stokes switching parameter that varies rapidly near the Stokes curve and is constant outside of the rapidly varying neighborhood. Inserting~\eqref{4:WKB} into~\eqref{e:R_N1} yields
\begin{align}\label{form}
	 -2A'\chi'\e^{-\chi/\eta}\sim-\frac{\eta^{2N-1}(\chi')^2\Gamma(2N+3/2)}{\chi^{2N+3/2}}\,.
\end{align}
We transform \eqref{form} to treat $\chi$ as the independent variable and write $\chi= r\e^{i\theta}$ in polar coordinates. We fix $r$, and we consider only variations
in the angular direction. Using the optimal truncation $N=N_{\mathrm{opt}}$ and simplifying using Stirling's formula gives
\begin{align}
	\frac{\d A}{\d\theta}\sim \frac{\i}{\eta^2}\sqrt{\frac{\pi \rho}{2}}\exp\left(\frac{\rho}{\eta}\left(\e^{\i\theta}-1\right)-\frac{\i\theta r}{\eta}-\i\theta(1/2+2\omega)\right)\,.
\label{e:A}
\end{align}
The right-hand side of~\eqref{e:A} is exponentially small in $\eta$, except in the neighborhood of $\theta=0$. Defining an inner region
$\theta=\eta^{1/2}\bar{\theta}$, we find that
\begin{align} \label{inter}
	\frac{\d A}{\d\bar{\theta}}\sim\i\sqrt{\frac{\pi \rho}{2\eta^3}}\e^{-r\bar{\theta}^2/2}\,.
\end{align}
By integrating \eqref{inter}, we see that the behavior of $A$ as the Stokes curve is crossed is
\begin{align}\label{inter2}
	A\sim \i \sqrt{\frac{\pi }{2\eta^3}}\int_{-\infty}^{\sqrt{\rho}\bar{\theta}}\e^{-s^2/2}\d s\,.
\end{align}
\textcolor{black}{}Evaluating the integral in \eqref{inter2} and using the form of $R_N$ from~\eqref{4:WKB}, we find that the exponentially small contribution from $\xi_{1-}$ is
\begin{equation}
	[R_N]_-^+ \sim \frac{\i\pi\Lambda}{ \eta^{3/2}\sqrt{\chi'}} \mathrm{e}^{-\chi/\eta}\,.
\label{e:diatomic_RN}
\end{equation}
\textcolor{black}{}To capture the change across the Stokes curve, we also need to include the contribution from $\xi_{0-}$ that is conjugate to $\xi_{1-}$; this contribution is given by the complex conjugate of~\eqref{e:diatomic_RN}. Adding~\eqref{e:diatomic_RN} and its complex conjugate, we find that the change in the exponentially small contribution as the Stokes curve is crossed from left to right is
\begin{equation}
	[R_N]_-^+ \sim \frac{\i\pi\Lambda}{ \eta^{3/2}\sqrt{\chi'}} \mathrm{e}^{-\chi/\eta} + \mathrm{c.c.}\,,
\end{equation}
where the complex conjugate c.c. indicates
the contribution from the singularity at $\xi_{0-}$.

The overall exponentially small contribution to the asymptotic behavior of $v(\xi)$ in the wake of the leading-order solitary wave is given by the sum of the contributions from each of the four singularities (see Fig.~\ref{f:singularity}).
In the limit $\eta \rightarrow 0$, the exponentially small terms have the asymptotic expression
 \begin{align} \nonumber
	v_{\exp}\sim
\frac{\i\pi}{\eta^{3/2}}\e^{-\mathrm{Re}(\chi_{1-}(\xi))/\eta}
&\frac{\Lambda_{1-}\e^{-\i\mathrm{Im}(\chi_{1-}(\xi))/\eta}}{\sqrt{\chi_{1-}'(\xi)}}\\
	+&\frac{\i\pi}{\eta^{3/2}}\e^{-\mathrm{Re}(\chi_{1+}(\xi))/\eta}
\frac{\Lambda_{1+}\e^{-\i\mathrm{Im}(\chi_{1+}(\xi))/\eta}}{\sqrt{\chi_{1+}'(\xi)}}+\mathrm{c.c}\,,
\label{e:4singularities}
\end{align}
where $\chi_{1-}$ is the singulant that is associated with $\xi=\xi_{1-}$ and $\chi_{1+}$ is the singulant that is associated with $\xi=\xi_{1+}$. Using the relation $\mathrm{Re}(\chi_{1+})=\mathrm{Re}(\chi_{1-})$ from~\eqref{e:chiintrealimag}, the asymptotic expression~\eqref{e:4singularities} becomes
\begin{equation}
	v_{\exp}\sim
\frac{\i\pi}{\eta^{3/2}}\e^{-\mathrm{Re}(\chi_{1-}(\xi))/\eta}
\left[\frac{\Lambda_{1-}\e^{-\i\mathrm{Im}(\chi_{1-}(\xi))/\eta}}{\sqrt{\chi_{1-}'(\xi)}}
	 +\frac{\Lambda_{1+}\e^{-\i\mathrm{Im}(\chi_{1+}(\xi))/\eta}}{\sqrt{\chi_{1+}'(\xi)}}\right]+\mathrm{c.c}\,.
\end{equation}
In Appendix \ref{app:local1}, we show that $\Lambda_{1+} = \i \Lambda_{1-}$. Far behind the central solitary wave, $|u_0(\xi-1)-u_0(\xi+1)|\to0$ because the leading-order traveling wave is exponentially localized. Consequently, $\chi_{1,\pm}'(\xi)\sim \i\sqrt{3}\delta^{1/4}/c_{\epsilon}$ in the limit $\xi \rightarrow -\infty$, as indicated in \eqref{e:chisquare}. This yields a convenient trigonometric simplification:
\begin{align}
	\nonumber v_{\exp}\sim\frac{4\pi\Lambda_{1-}}{\eta^{3/2}\sqrt{\chi_{1-}'(\xi)}}\e^{-\mathrm{Re}(\chi_{1-}(\xi))/\eta}
\cos\bigg(\frac{\mathrm{Im}(\chi_{1-}(\xi)-\chi_{1+}(\xi))}{2\eta}&+\frac{\pi}{4}\bigg)\\
\times\cos&\bigg(\frac{\mathrm{Im}(\chi_{1+}(\xi)+\chi_{1-}(\xi))}{2\eta}-\frac{\pi}{4}\bigg)\,.
\label{e:diatomic_oscillation}
\end{align}
From the integral expression~\eqref{e:chiint} for the singulant $\chi$, we see that $\mathrm{Re}(\chi_{1,\pm})$ and $\mathrm{Im}(\chi_{1-})-\mathrm{Im}(\chi_{1+})$ do not depend on $\xi$. Consequently, we
write the amplitude of the far-field waves as
\begin{equation}
	v_{\mathrm{amp}}\sim\frac{4\pi c_{\epsilon}^{1/2}\Lambda_{1-}}{  3^{1/4}\delta^{1/8} \eta^{3/2}}\e^{-\mathrm{Re}(\chi_{1-}(\xi))/\eta}
\cos\bigg(\frac{\mathrm{Im}(\chi_{1-}(\xi)-\chi_{1+}(\xi))}{2\eta}+\frac{\pi}{4}\bigg) \quad \mathrm{as} \quad \eta \rightarrow 0\,.
\label{e:diatomicamp}
\end{equation}

%%%%

\subsection{Comparison of our Asymptotic and Computational Results}\label{s:dimernum}

As in Section~\ref{s:woodnum}, we compare our asymptotic results for the far-field amplitude to numerical simulations. We again use the velocity Verlet algorithm and again simulate the equations of motion on a large periodic domain with relative coordinates. For diatomic Hertzian chains, the relative coordinate system is given in terms of $r(n,t)$, where
\begin{align}
	r(2n,t)=v(2n+1,t)-u(2n,t)\,, \quad r(2n+1,t)=u(2n+2,t)-v(2n+1,t)\,.
\end{align}
In this coordinate system, the equations of motion \eqref{e:diatomic1}--\eqref{e:diatomic2} become
\begin{align}
	\nonumber \ddot{r}(2n,t)&=\bigg(1+\frac{1}{\eta^2}\bigg)[\delta-r(2n,t)]^{3/2}_+
                           -[\delta-r(2n-1,t)]^{3/2}_+-\frac{1}{\eta^2}[\delta-r(2n+1,t)]^{3/2}_+\,,\\
	\nonumber \ddot{r}(2n+1,t)&=\bigg(1+\frac{1}{\eta^2}\bigg)[\delta-r(2n+1,t)]^{3/2}_+
                           -\frac{1}{\eta^2}[\delta-r(2n,t)]^{3/2}_+-[\delta-r(2n+2,t)]^{3/2}_+\,.
\end{align}
As with the woordpile chain, the domain includes $M=2^{10}$ particles, with indices $n \in \{-M/2+1, \ldots, M/2\}$.
{At each time step, we multiply the solution by the windowing function in \eqref{e:window}.} The initial condition is given by the leading-order solution~\eqref{e:leading}.

\begin{figure}[tb]
\centering

\subfloat[$\delta = 3$]{
\includegraphics{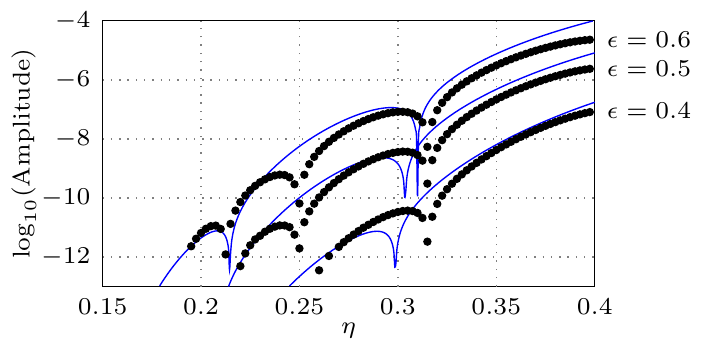}
}
\subfloat[$\delta = 5$]{
\includegraphics{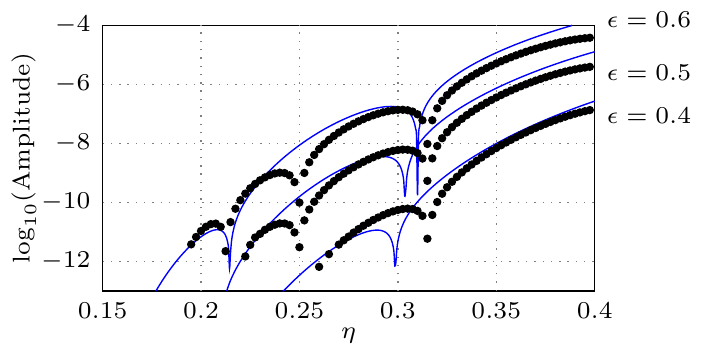}
}

\subfloat[$\delta = 10$]{
\includegraphics{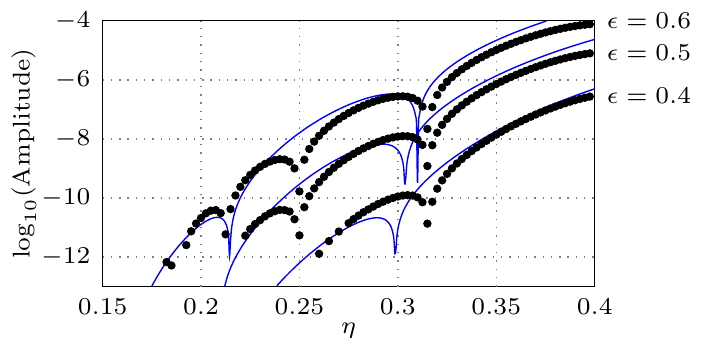}
}
\subfloat[$\delta = 15$]{
\includegraphics{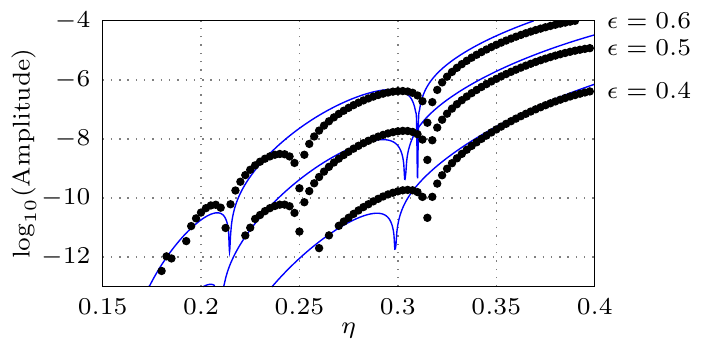}
}

\caption{Comparison of our asymptotic results and numerical simulations of the far-field wave amplitude of nanoptera in a diatomic Hertzian chain with precompressions of (a) $\delta = 3$, (b) $\delta = 5$, (c) $\delta = 10$, and (d) $\delta = 15$.
 For each value of $\delta$, we show our computations for long-wavelength parameters of $\epsilon = 0.4$, $\epsilon = 0.5$, and $\epsilon = 0.6$. Our asymptotic approximation
 of the far-field amplitude is accurate for a wide range of values of the mass-ratio parameter $\eta$, with certain important exceptions. For fixed values of $\delta$ and $\epsilon$, two oscillatory wave trains interfere destructively for specific values of $\eta$.
 Our asymptotic approximation is capable of predicting the destructive interference that occurs for $\eta \approx 0.32$ and $\eta \approx 0.22$, and the accuracy of our 
 approximation becomes progressively better for progressively smaller values of $\eta$. However, the approximation does not predict the destructive interference that occurs for $\eta \approx 0.25$ and $\eta \approx 0.19$. Our approximation for the {anti-resonance conditions} becomes less accurate for progressively smaller $\epsilon$.
}\label{F:Diatomic}
\end{figure}

We calculate the amplitude of the far-field waves for precompressions of $\delta = 5$, $\delta = 7.5$, $\delta = 10$, and $\delta = 15$ with long-wavelength parameters of $\eps = 0.4$, $\eps = 0.5$, and $\eps = 0.6$ for a range of values of the mass-ratio parameter $\eta$. As with woodpile chains, we run each simulation for a sufficiently long time so that the amplitude of the far-field oscillations {appears to
reach a constant value.} In Fig.~\ref{F:Diatomic}, we present the results of our computations and compare them to our asymptotic results from \eqref{e:diatomicamp}.

We make two observations from Fig.~\ref{F:Diatomic}. First, our asymptotic approximation accurately predicts the amplitude of the far-field waves for a wide
range of values of $\eta$. Second, our asymptotic calculation does not always identify the values of $\eta$ for which destructive interference causes the far-field oscillations to vanish. In each panel of Fig.~\ref{F:Diatomic}, we observe four values of $\eta$ (these are $\eta \approx 0.32$, $\eta \approx 0.25$, $\eta \approx 0.22$, and $\eta \approx 0.19$) at which the far-field waves in our numerical computations vanish, corresponding to localized solitary waves with no {exponentially small oscillations behind the leading-order solitary wave}.
From \eqref{e:diatomicamp},
our asymptotic analysis predicts these values to be solutions of
\begin{equation}
	 \cos\bigg(\frac{\mathrm{Im}(\chi_{1-}(\xi)-\chi_{1+}(\xi))}{2\eta}+\frac{\pi}{4}\bigg) = 0\,,
	 \label{e:ortho0}
\end{equation}
which simplifies to solving
\begin{align} \label{e:ortho}
	2\pi\eta \left(K+\frac{1}{4}\right)=\frac{\sqrt{3}\i}{c_\epsilon}\int_{\xi_{1-}}^{\xi_{1+}}\bigg(\delta +\frac{(u_0(\xi-1) - u_0(\xi+1))}{2}\bigg)^{1/4}\d s
\end{align}
for integer values of $K$. The expression \eqref{e:ortho} is an orthogonality condition \cite{Lustri,Lustri1,Vainchtein,Jayaprakash,Alfimov}. Values of $\eta$ that satisfy \eqref{e:ortho} correspond to dips in the asymptotically predicted amplitude in Fig.~\ref{F:Diatomic}. These values yield localized solitary waves. We see that the condition \eqref{e:ortho} is capable of predicting the localized solitary waves that arise for $\eta \approx 0.32$ and $\eta \approx 0.22$. However, it fails to predict the localized solitary waves that arise for $\eta \approx 0.25$ and $\eta \approx 0.19$. In fact, our exponential asymptotic analysis appears to systematically miss every second point (i.e., if it captures one point, then it misses the next point, and vice versa) at which the oscillations vanish. Additionally, for progressively smaller values of $\epsilon$, the orthogonality condition \eqref{e:ortho} becomes progressively less accurate at predicting the values of $\eta$ that give localized solitary waves.

Systematically missing half of these solitary-wave solutions appears to be a limitation of the method that we used to obtain our asymptotic expressions.
There are several approximations in our analysis. Notably, we analytically continued the long-wavelength KdV approximation from the real axis into the complex plane. The orthogonality condition \eqref{e:ortho} depends on the behavior of $u_0(\xi)$ on a contour that connects the singularities at $\xi_{1-}$ and $\xi_{1+}$ (see Fig.~\ref{f:singularity}). These singularities are far from the real axis, and their distance from the real axis increases as the long-wavelength parameter $\epsilon$ decreases. Consequently, it is possible that errors in this approximation --- and, in particular, in the locations of the singular points --- become progressively larger for progressively smaller $\epsilon$.

Detecting destructive interference involves determining configurations for which two oscillations with very short wavelengths precisely cancel each other. The individual Stokes contributions have wavelengths that are of size $\mathcal{O}(\eta)$ as $\eta \rightarrow 0$. Errors in the positions of the singularities on this very short scale are capable of significantly changing the values of $\eta$ for which cancellation occurs. It is possible that approximation errors in the leading-order solution lead to shifts in the singularities in the analytically-continued solution.
This may disrupt the precise cancellation that is necessary for the manifestation of localized solitary-wave solutions, causing our asymptotic approximation to miss some of them.

The issue of missing solitary-wave solutions did not arise in a recent study of a diatomic FPUT chain \cite{Lustri1}, despite the use of a similar long-wavelength approximation to continue a leading-order solution into the complex plane. It is not apparent why the method worked effectively for diatomic FPUT chains but has failed to identify analogous behavior in our diatomic Hertzian chain. This is an interesting open question that merits further study.

%%%%

\section{Conclusions and Discussion}
\label{s:conclusion}

In the present study, we derived asymptotic approximations for traveling waves in two particle chains that are singular perturbations of a monoatomic particle chain with precompression. Specifically, we considered a woodpile chain and a diatomic Hertzian chain with a small ratio between the masses of its two types of particles. {The traveling-wave solutions of these systems are nanoptera, which consist of solitary wave along with
non-decaying oscillations of exponentially small amplitude.}

{A monoatomic Hertzian chain supports traveling solitary-wave solutions, which one can approximate as soliton solutions of the KdV equation under suitable assumptions.} We used this monoatomic solitary wave as the leading-order behavior in both woodpile chains and diatomic Hertzian chains. In both systems, we found that non-decaying oscillations appear in the wake of a primary traveling wave. These oscillations arise from the Stokes phenomenon, in which exponentially small contributions to an analytically-continued solution switch on or off as Stokes curves are crossed in the complex plane.

By employing exponential asymptotic analysis, we obtained an asymptotic form for these exponentially small contributions, and we used this form to determine the amplitude of the non-decaying far-field waves that appear behind the leading-order wave. For woodpile chains, the solution has one Stokes curve, which produces one exponentially small oscillation in the region behind the wave front. We determined an asymptotic approximation of this behavior, and we compared our asymptotic results with numerical computations. We saw that the asymptotic prediction accurately approximates the wave behavior in the far field. However, the asymptotic approximation of the far-field wave amplitude becomes progressively less accurate for progressively smaller values of the precompression parameter $\delta$. 
{It is possible that this is a consequence of the inadequacy of examining a weakly nonlinear regime of the woodpile chain as nonlinear effects become more significant, leading to errors in the description of exponentially small effects. It would be interesting to study woodpile chains without
precompression to investigate whether it is possible to analyze the behavior of the far-field oscillations in the strongly nonlinear regime of woodpile systems using exponential asymptotic methods.} We discuss the feasibility of such a study at the end of this section.

We also used exponential asymptotic analysis to study nanoptera in a diatomic Hertzian chain in which the mass ratio between light and heavy particles is small.
We found that there are two Stokes curves in the analytically-continued traveling-wave solution. These produce two exponentially small oscillations --- which have identical amplitude but different phases --- behind the wave front. There exists a set of values of the mass-ratio parameter $\eta$ at which the oscillations interfere destructively to produce a localized solitary wave.

Our comparison of the asymptotic results and numerical computations in diatomic Hertzian chains shows that our exponential asymptotic analysis accurately approximates
the amplitude of the far-field waves for a wide range of values of $\eta$. However, our computations also reveal that our asymptotic analysis is unable to detect the wave cancellation that occurs for several values of $\eta$. Furthermore, the predicted mass ratio values that produce wave cancellation
{are less accurate than those in analyses of other particle chains (such as the exponential asymptotic calculations from \cite{Lustri,Lustri1}), particularly for small values of the long-wavelength parameter $\epsilon$.} It is possible that this inaccuracy arises from our use of a long-wavelength approximation for the leading-order solution; this approximation {may deviate from the exact solution away from the real axis.}
This is an important question, as the Stokes phenomenon depends on the behavior of singularities in the analytically-continued solution. {It is worthwhile to study whether using an alternative method of approximating the leading-order solution can capture all of the values of $\eta$ that {cause the oscillations to cancel}.}

Our work also opens {the following} interesting questions.
In the present study, we used a similar asymptotic approach as the one in \cite{Lustri,Lustri1} and evaluated our asymptotic approximations by comparing them to numerical computations. It would be interesting to develop a rigorous existence proof of nanopteron solutions in both diatomic Hertzian chains and woodpile chains. One possible approach for attempting this is
by adapting the Beale-ansatz method of \cite{Faver,Hoffman}.

As we discussed in Section \ref{intro},
one-sided nanopteron solutions in both woodpile and diatomic Hertzian chains are metastable because the oscillations slowly draw energy away from the wave front. This indicates that the wave cannot persist indefinitely; instead, it must eventually decay. This property is not present in the leading-order approximation, nor it is apparent in the form of the exponentially small oscillations in the analysis either in the present study or in~\cite{Lustri,Lustri1}. The long-time decay must arise from interactions between the leading-order traveling wave and the exponentially small oscillations. It would be useful to determine (1) how the leading-order wave and the exponentially small oscillations interact and (2) whether or not one can detect such interactions by using an appropriate long-time rescaling or by calculating higher-order corrections to the oscillatory wave train.

In our analysis, we considered a weakly nonlinear regime for both woodpile and diatomic Hertzian chains. In principle, one can apply the exponential asymptotic method that was developed in \cite{Chapman} with a few additional complications to study strongly nonlinear systems, such as a singularly perturbed Hertzian system without precompression. However, there are two significant challenges that need to be overcome to undertake such an analysis. The first challenge is that one no longer has access to a closed-form leading-order solution to analytically continue in a straightforward fashion.
Perhaps it is possible to use the leading-order approximation that was derived in \cite{sen:2008},
but it is not apparent whether this approximation is valid if it is analytically continued away from the real axis. This is an important point to clarify because the Stokes phenomenon is caused by singularities in the complex plane. The second challenge relates to the form of the potential in~\eqref{e:potential}. This potential is not smooth at any point at which the argument in the brackets is $0$. In the present study, the precompression parameter $\delta_0$ is
larger than or equal to
the relative displacement of two adjacent particles, so the argument in the brackets is always non-negative and the potential in~\eqref{e:potential} is smooth.
Without precompression, however, the argument in the brackets can switch between positive and negative signs, and the potential is not smooth everywhere.
An important direction for future studies is to explore the effects of such a non-smooth potential on exponential asymptotic analysis. If these challenges are overcome, it would be useful to compare the ensuing results with {those of} {the uncompressed woodpile~\cite{Xu} and diatomic Hertzian~\cite{Jayaprakash} chains.}

%%%%%%%%%
\section{Acknowledgement}
CJL and GD were supported by Australian Research Council Discovery Project DP190101190.
%%%%%%%%%
\appendix

\section{Determining the Prefactor Constants}

We determine the prefactor constants for both the woodpile chain and the diatomic Hertzian chain.
{
Assuming that the power series diverges in a factorial-over-power fashion, it must cease to be asymptotic in a narrow region in the neighborhood of the singularities in the complex plane. Inside this region, the earlier terms in the series are not larger asymptotically than the later series terms in the limit $\eta \rightarrow 0$.
Describing the solution behavior near one of these singularities requires obtaining a local expansion of the solution near the singular point. Once we have obtained this local expansion, we calculate the prefactor constants in the late-order terms using asymptotic matching to ensure that the local expansions are consistent with the power-series behavior.}

%%%%

\subsection{Woodpile Chain}
\label{app:localw}

To determine the value of $\Lambda_+$, we match the late-order expansion in the outer region with the local solution in an inner region near the singularity at $\xi = \xi_{0+}$ by using Van Dyke's matching principle.

As $\xi \rightarrow \xi_{0+}$, we find that
\begin{align}
	u_0(\xi) &\sim -\frac{1}{\alpha-1}\frac{\delta }{\xi - \xi_{0+}} + \mathcal{O}(\xi - \xi_{0+})\,, \quad & u_0(\xi + 1) &\sim \frac{1}{\alpha-1}\delta \epsilon\, \mathrm{coth}(\eps) + \mathcal{O}(\xi - \xi_{0+})\,,\\
	v_0(\xi) &\sim -\frac{1}{\alpha-1}\frac{\delta }{\xi - \xi_{0+}} + \mathcal{O}(\xi - \xi_{0+})\,, \quad & u_0(\xi - 1) &\sim -\frac{1}{\alpha-1}\delta \epsilon \,\mathrm{coth}(\eps) + \mathcal{O}(\xi - \xi_{0+})\,.
\end{align}

To locate the relevant inner region, we need to determine where the validity of the late-order term ansatz breaks down. From the form of the late-order ansatz~\eqref{2:ansatz}, we see that this occurs for $\eta^2 \chi^{-2} = \mathcal{O}(1)$ as $\eta \rightarrow 0$. That is, it occurs for $\eta^2(\xi-\xi_{0+})^{-2} = \mathcal{O}(1)$; this corresponds to the inner scaling $\xi - \xi_{0+} = \eta \overline{\xi}$. From asymptotic balancing, the appropriate rescaled inner variables are
\begin{equation}\label{2:rescaling}
	u(\xi) = -\frac{1}{\alpha-1}\frac{\delta  }{ \eta\overline{\xi}} + \hat{u}(\overline{\xi})\,,\quad u(\xi+1) =\hat{u}(\overline{\xi }+ \eta^{-1})\,, \quad u(\xi-1) =\hat{u}(\overline{\xi}-\eta^{-1})\,, \quad v(\xi) = -\frac{1}{\alpha-1}\frac{\delta  }{\eta \overline{\xi}}+ \frac{\hat{v}(\overline{\xi})}{\eta}\,.
\end{equation}
Retaining the leading-order terms as $\eta \rightarrow 0$, the rescaled inner equation gives
\begin{equation}
	-\frac{1}{\alpha-1}\frac{2\delta }{\overline{\xi}^3} + \diff{^2\hat{v}(\overline{\xi})}{\overline{\xi}^2}
	= -\frac{k}{c_{\epsilon}^2} \hat{v}(\overline{\xi})\,.
\label{e:appendix_rescaled}
\end{equation}
We express $\hat{v}$ as a power series
\begin{equation}
	\hat{v}(\overline{\xi}) \sim \sum_{j=1}^{\infty}\frac{v_n}{\overline{\xi}^{2j+1}}\quad \mathrm{as} \quad \overline{\xi} \rightarrow 0\,,
\label{e:appendix_inner}
\end{equation}
and we note that we include the leading-order singularity as part of the rescaling process \eqref{2:rescaling}. Substituting~\eqref{e:appendix_inner} into~\eqref{e:appendix_rescaled} yields
\begin{equation}
	-\frac{1}{\alpha-1}\frac{2\delta }{\overline{\xi}^3} +\sum_{j=1}^{\infty}\frac{(2j+1)(2j+2)v_j}{\overline{\xi}^{2j+3}} = -\frac{k}{c_{\epsilon}^2}  \sum_{j=1}^{\infty}\frac{v_j}{\overline{\xi}^{2j+1}}\,.
\end{equation}
By matching orders of $\overline{\xi}$, we obtain the recurrence relation
\begin{equation}\label{recur}
	v_1 = 2\delta c_{\epsilon}^2/(k(\alpha-1)) \,, \quad (2j+2)(2j+1)c_{\epsilon}^2 v_{j} = - k v_{j+1}\,.
\end{equation}
Solving the recurrence relation \eqref{recur} gives
\begin{equation}\label{series_app}
	v_j = \frac1{\alpha-1}\delta
     \left(-1\right)^{j+1}\left(\frac{ c_{\epsilon}^{2}}{k}\right)^{j}\Gamma(2j+1)\,.
\end{equation}
By comparing the series expression \eqref{series_app} with the inner limit of the late-order ansatz, we obtain
\begin{equation}\label{A:Lambda}
	\Lambda_+ = \lim_{j\rightarrow \infty} \frac{v_j (\i \sqrt{k}/c_{\epsilon})^{2j+1}}{\Gamma(2j+1)}
	        = -\frac{1}{\alpha-1}\frac{\i\delta  \sqrt{k}}{c_{\epsilon}}\,.
\end{equation}

%%%%%%

\subsection{Diatomic Hertzian Chain}
\label{app:local1}

To determine $\Lambda_{1-}$, we match the late-order term expansion in the outer region with a local expansion in an inner region near $\xi=\xi_{1-}$. We perform this analysis in the neighborhood of $\xi=\xi_{1-}$. From~\eqref{e:lateorder1}, we see that the factorial-over-power ansatz breaks down when $\eta^2\chi^{-2}=\mathcal{O}(1)$ as $\eta\to0$. We introduce an inner scaling $\xi-\xi_{1-}=\eta^{4/3}\bar{\xi}$, which gives the
rescaled inner variables
\begin{align}
	u(\xi+1)=-\frac{8\delta}{\eta^{4/3}\bar{\xi}}+\hat{u}(\bar{\xi}+\eta^{-4/3})\,,
\quad  u(\xi-1)=\hat{u}(\bar{\xi}-\eta^{-4/3})\,,
	\quad v(\xi)=-\frac{4\delta}{\eta^{4/3}\bar{\xi}}+\frac{\hat{v}(\bar{\xi})}{\eta^{4/3}}\,.
\label{e:UVinner}
\end{align}
Retaining the leading-order terms as $\eta\to0$, the rescaled inner equation gives
\begin{align}
	-\frac{8\delta}{\bar{\xi}^3}+\frac{\d^2\hat{v}(\bar{\xi})}{\d{\bar{\xi}}^2}=
\frac1{c_\epsilon^2}
\left[\bigg(\frac{4\delta}{\bar\xi}-\hat{v}(\xi)\bigg)^{3/2}
	-\bigg(\frac{4\delta}{\bar\xi}+\hat{v}(\xi)\bigg)^{3/2}\right]\,.
\label{e:innerequation}
\end{align}
We express $\hat{v}$ in terms of a power series
\begin{align}
	\hat{v}(\bar{\xi})\sim\sum_{j=1}^{\infty}\frac{a_j}{\bar{\xi}^{\frac{3j}{2}+1}} \quad \mathrm{as}\quad \bar{\xi}\to0\,.
\label{e:v_inner}
\end{align}
Matching the inner expansion~\eqref{e:v_inner} with the outer ansatz~\eqref{e:v_outer} yields
\begin{align}
	\Lambda_{1-} = \lim_{j\to\infty}\frac{a_j\i^{2j+2}2^{5j+4}\delta^{j/2+1/2}}{3^{j+1/2}c_\epsilon^{2j+2}\Gamma(2j+3/2)}\,.
\label{e:Lambda}
\end{align}
We numerically compute values for $a_j$ by substituting the series expression~\eqref{e:v_inner} into~\eqref{e:innerequation} and solving a recurrence relation for $a_j$ in terms of the previous series coefficients. We obtain an approximation $\Lambda_{\mathrm{approx}}$ for $\Lambda_{1-}$ by evaluating the right-hand side of~\eqref{e:Lambda} for large values of $j$. When we take $j$ to be sufficiently large, we obtain an accurate approximation of the exact value of $\Lambda_{1-}$.

\begin{figure}
\centering
\includegraphics{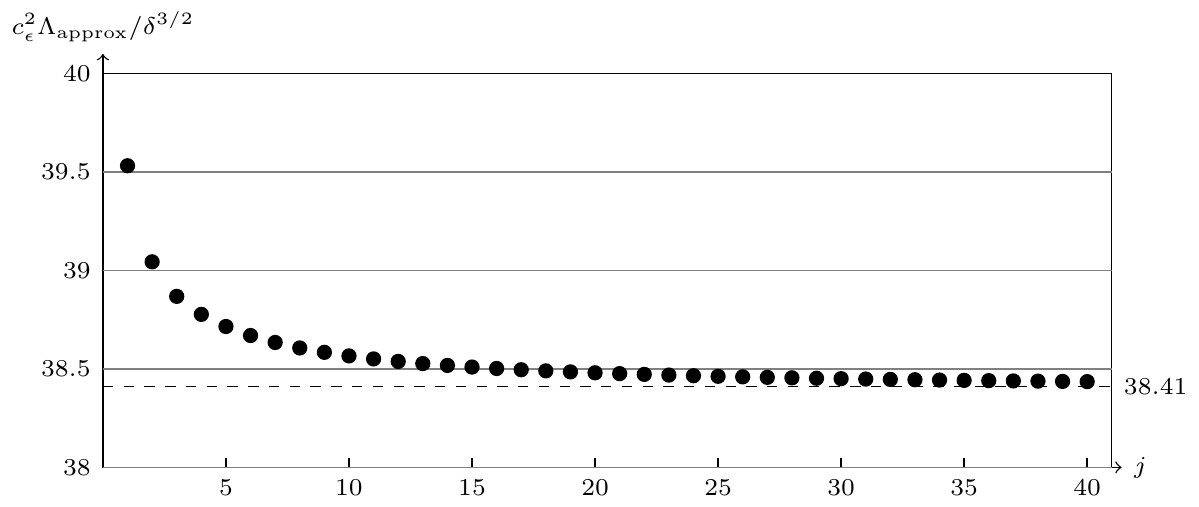}
\caption{Determining $\Lambda_{\mathrm{approx}}$ using ~\eqref{e:Lambda} for progressively larger values of $j$. The dashed line indicates the value to which the rescaled $\Lambda_{\mathrm{approx}}$ appears to converge as we increase $j$.}
\label{f:Lambda}
\end{figure}

In Fig.~\ref{f:Lambda}, we show the behavior of $\Lambda_{\mathrm{approx}}$ as we increase $j$; we observe that our approximation is converging. For sufficiently large values of $j$, we
compute that
\begin{equation}
	\Lambda_{1-}=38.41\, \delta^{3/2}/c_\epsilon^2\,.
\end{equation}

Following the same approach as above yields
\begin{equation}
	\Lambda_{1+}=\i\Lambda_{1-}\,.
\end{equation}

%%%%%%%%%%%%%%%%%%%%%%%%%%%%%%%%%%%%%%%%%%%%%%%%%%%%%%%%%%%%%%%%%%%%%%%%%%%%%%%%%%%%%%%%%%%%%%%%%%%%%%%%%%%%%%%%%%%%%%%%%%%%%%%%%%%%%%%\

\bibliography{reference4.bib}
\bibliographystyle{plain}

\end{document}